\documentclass[aps,prb,floatfix,superscriptaddress,twocolumn,hyperref=pdftex]{revtex4-2}

\usepackage{amssymb}
\usepackage{hyperref}
\usepackage{graphicx}
\usepackage{placeins}
\usepackage{booktabs}
\usepackage{svg}

\usepackage{dsfont}
\usepackage[utf8]{inputenc}
\usepackage[normalem]{ulem}

\usepackage{upgreek}
\usepackage{siunitx}
\usepackage{physics}
\usepackage{soul}
\usepackage{xcolor}
\usepackage{tabularx,multirow,array}
\usepackage{siunitx}
\usepackage{chemmacros}
\usepackage{cancel}
\usepackage{amsmath}

\usepackage{bbm}

\begin{document}

\title{Phase-controlled minimal Kitaev chain in multiterminal Josephson junctions}

\author{Samuel D. Escribano}
\email{samuel.diazes@gmail.com}
\affiliation{Department of Condensed Matter Physics, Weizmann Institute of Science, Rehovot 7610001, Israel}

\author{Anders Enevold Dahl}
\affiliation{Center for Quantum Devices, Niels Bohr Institute, University of Copenhagen, DK-2100 Copenhagen, Denmark}

\author{Karsten Flensberg}
\affiliation{Center for Quantum Devices, Niels Bohr Institute, University of Copenhagen, DK-2100 Copenhagen, Denmark}

\author{Yuval Oreg}
\affiliation{Department of Condensed Matter Physics, Weizmann Institute of Science, Rehovot 7610001, Israel}

\begin{abstract}
We propose a nanodevice based on multi-terminal Josephson junctions for the creation and detection of poor man’s Majorana (PMM) modes. The device consists of a double three-terminal Josephson junction (3TJJ) embedded in a planar semiconductor that engineers a two-site Kitaev chain. We identify the conditions necessary for the emergence of PMM modes by precisely tuning the superconducting phases of the terminals and the inter-junction coupling between the 3TJJs. To validate our findings, we perform numerical simulations that account for disorder in a more general geometry and test experimentally feasible protocols based on conductance measurements to reliably detect these modes. Unlike traditional approaches that rely on magnetic fields, our design eliminates the need for such fields, potentially increasing the energy-level resolution and significantly expanding the range of compatible semiconductor materials, such as Ge-based heterostructures.
\end{abstract}

\maketitle

\section{Introduction}
Majorana modes are of fundamental interest because of their non-Abelian statistics and potential for fault-tolerant quantum operations~\cite{Alicea:IOP12, Aguado:rnc17, Marra:JOP22}. Recently, the bottom-up approach of engineering Kitaev chains~\cite{Kitaev:PU01} using fine-tuned quantum dots (QDs)~\cite{Leijnse:PRB12, Sau:NatCom12, Fulga:IOP13, Dvir:Nat23, Haaf:Nat24, Tsintzis:PRXQ24, Zatelli:NatCom24, Bordin:arxvi24, Souto:arxiv24} has demonstrated significant advantages over traditional top-down methods~\cite{Lutchyn:NRM18, Prada:NRP20, Flensberg:NRM21}. This is primarily due to their superior tunability of system parameters, which enable the detection and control of Majorana bound states (MBSs) with greater confidence and precision.

\begin{figure}[h!]
\begin{centering}
\includegraphics[width=0.99\columnwidth]{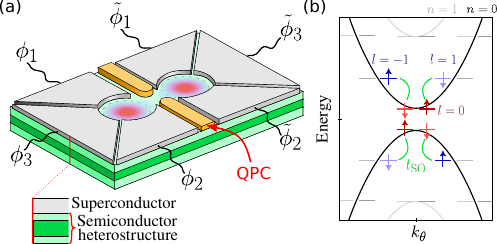}
\par\end{centering}
\caption{\textbf{\label{fig1} System sketch. (a)} Schematic of the studied device: a double three-terminal Josephson junction (3TJJ) coupled via a quantum point contact (QPC). Each 3TJJ consists of three superconductors (SCs), in gray, each characterized by a different superconducting phase: $\phi_1$, $\phi_2$, and $\phi_3$ for the left junction; and $\tilde\phi_1$, $\tilde \phi_2$, and $\tilde \phi_3$ for the right junction. The three SCs are grown on top of a planar semiconductor (SM) heterostructure, in green (active layer in dark green), and converge into an uncapped circular tunneling region, whose transparency can be tuned using a top and/or a bottom gate. This creates some sort of confinement in this area, represented by a red-shaded area. The QPC controls the tunneling between both junction regions. \textbf{(b)} Energy level diagram for each tunneling region. We assume harmonic quantum confinement within each tunneling region, which transforms the parabolic dispersion into discreet levels. Without spin-orbit~(SO) interaction, the energy levels are labeled by their angular momentum quantum number $l$. SO interaction couples states with opposite spins and $l\pm 1$ angular momenta through $t_{\rm SO}$. The colored levels represent the lowest-energy states that span an independent Hilbert subspace. Superconductivity further couples electron (upside parabola) and hole (downside parabola) levels with opposite spin (coupling not represented).}
\end{figure}

Current strategies for realizing MBSs often rely on the application of magnetic fields and the use of QDs that are weakly~\cite{Dvir:Nat23, Haaf:Nat24} or moderately~\cite{Zatelli:NatCom24} coupled to superconductors~(SCs). However, these approaches have several drawbacks. The presence of a magnetic field introduces pair-breaking effects which lead to small minigaps and broadened energy levels, likely to limit the detection and measurement of the MBS's lifetimes. Additionally, the weak-coupling with the SCs exacerbate charge fluctuations within the system, further compromising its stability. The need for magnetic fields also imposes stringent material requirements, limiting the choice of semiconductor~(SM) platforms and the compatible SCs that can be grown epitaxially. Such systems require SMs with both strong $g$-factors and robust SO coupling~\cite{Oreg:PRL10, Lutchyn:PRL10, Klinovaja:PRB12}. As a result, materials such as Ge-based heterostructures, which lack sufficiently strong $g$-factors~\cite{Giorgioni:NatCom16, Scappucci:NRM21}, are less suitable for these platforms even though they exhibit quite strong SO couplings~\cite{Hao:NanoLet10, Higginbotham:PRL14, Wang:IOP17}.

In this work, we propose an alternative nanodevice for creating bottom-up Kitaev chains, based on multi-terminal Josephson junctions. This device requires only moderately strong-coupling to SCs and completely eliminates the need for magnetic fields, significantly broadening the range of compatible SM materials, including III-V compounds and Ge-based heterostructures. Multi-terminal Josephson junctions have been shown to break time-reversal symmetry analogously to magnetic fields~\cite{Beri:PRB08, Heck:PRB14, Riwar:NatCom16, Coraiola:NatCom23, Coraiola:PRX24}, but without introducing pair-breaking decoherence effects. Moreover, several studies~\cite{Lesser:PRB21, Lesser:PNAS21, Lesser:PRB22, Luethi:PRB23} have proposed using SNSNS junctions for the creation of MBSs in top-down approaches.

Here, we demonstrate that poor man’s Majorana~(PMM) modes~\cite{Leijnse:PRB12} can be realized in a double three-terminal Josephson junction~(3TJJ) configuration, as illustrated in Fig.~\ref{fig1}(a), which corresponds to a two-site Kitaev chain. By precisely tuning the superconducting phases and the coupling between the two 3TJJs, we show that PMM modes can robustly emerge under experimentally accessible conditions. In fact, recent works have demonstrated that the relative phases between superconducting electrodes can be controlled with high precision using superconducting loops placed at a distance from the sample and in proximity to electric currents~\cite{Coraiola:NatCom23, Coraiola:PRX24}. Alternatively, it can also be achieved by injecting currents into the superconductors within complex geometries~\cite{Lykkegaard:arxiv24}, without introducing any current into the system itself. In both cases, it is noteworthy that the proposed nanodevice is entirely all-electric and does not require the application of magnetic fields. 

We begin in Sec.~\ref{Sec:1} by introducing a minimal model for a single 3TJJ, relating its parameters to material and geometric properties. Using this model, we identify the optimal conditions for observing zero-energy modes~(ZEM). We then extend our study to a double 3TJJ system in Sec.~\ref{Sec:2}, incorporating coupling through a quantum point contact (QPC). To evaluate the robustness and detectability of these modes, we numerically simulate the more general Hamiltonian under realistic conditions, including disorder and multimode effects. Our results show that PMM modes can be obtained under specific configurations of phase differences and coupling strengths. Importantly, in Sec.~\ref{Sec:3} we propose an experimental protocol based on local and non-local conductance measurements to obtain and detect the PMM modes. We argue moreover in Appendix~\ref{SM:Scalability} that our design is scalable, offering potential applications in more complex topological quantum architectures. The data to reproduce all the figures of this article are openly available in Ref.~\onlinecite{dataset}.

\section{Single three-terminal Long Josephson Junction}
\label{Sec:1}
We start by analyzing the case of an individual 3TJJ [see sketch on Fig.~\ref{fig2}(a)]. Although the general phenomenology of such a system has been discussed previously for long junctions~\cite{Heck:PRB14} (i.e., when the junction size is larger or comparable to the SO length), here we develop a different minimal model whose parameters are intimately related to the material and geometric properties that define the 3TJJ. Our goal is to validate this model against established results and identify optimal experimental conditions for observing MBSs.

\subsection{Model}
We describe the 3TJJ through the following Hamiltonian (see Appendix~\ref{SM:Model} for a detailed derivation)
\begin{equation}
    H(\vec{r},\omega) = H_{\rm 2DEG}(\vec{r}) +  \sum_k^{1,2,3} \Sigma_{\rm SC}^{(k)}(\vec{r},\omega),
\label{Eq:H_continuum}
\end{equation}
where
\begin{equation}
    H_{\rm 2DEG}(\vec{r})=\left(\frac{\hbar^2\vec{k}^2}{2m^*}+\frac{1}{2}m^*\omega_{\rm o}^2\left|\vec{r}\right|^2\right)\tau_z+\vec{\alpha}\cdot(\vec{\sigma}\times\vec{k})\tau_z,
\end{equation}
describes the 2DEG (or 2DHG) in which the tunneling region of the 3TJJ is electrostatically defined. Here, $m^*$ is the effective mass of the 2DEG and $\alpha$ the SO coupling. The potential in the tunneling region (second term), controlled by a top gate, is approximated as a harmonic oscillator potential with localization length  $l_\omega=\sqrt{\frac{\hbar}{m^*\omega_{\rm o}}}$, being $\omega_{\rm o}$ the frequency of the oscillator. The operators $\sigma_i$ and $\tau_i$ are the Pauli matrices in spin and Nambu space, respectively; while the vectors $\vec{k}$ and $\vec{r}$ stem for the momentum and position operators. In this model, we assume that the active layer of the SM heterostructure is thin enough that the subband spacing in the $z$ direction dominates the system's energy scales, allowing only a single subband to be occupied. Thus, this Hamiltonian is simplified to describe only this transverse subband, and the position vector $\vec{r}$ refers to the two in-plane coordinates, chosen here as the polar coordinates $\vec{r}=(r,\theta)$ (and similarly for $\vec{k}$). This assumption allows the model to remain general and applicable across materials such as the conduction band of InAs or the top valence band of~Ge. We choose the origin of coordinates at the center of the 3TJJ.

The second term in Eq.~\eqref{Eq:H_continuum} describes the self-energy due to each SC lead, labeled by $k$, which is obtained after integrating out the SC degrees of freedom
\begin{eqnarray}
	\Sigma_{\rm SC}^{(k)}(\vec{r},\omega) &=& -\frac{i\Gamma_{\rm N}(\vec{r})}{\sqrt{\Delta^2-\omega^2}}\left[\omega\tau_0 \right. \nonumber \\
	  &+&\left.\Delta\cos(\phi_k)\tau_x-\Delta\sin(\phi_k)\tau_y\right],
      \label{Eq:H_continuum2}
\end{eqnarray}
where $\Delta$ is the parent superconducting gap, $\omega$ the energy, and $\Gamma_{\rm N}$ is the coupling strength between the SC and SM. This coupling strength depends on experimental parameters and is uniform across points $\vec{r}$ directly beneath the SC regions, vanishing elsewhere.

From this continuum Hamiltonian we derive a minimal Hamiltonian to describe the lowest-energy modes. To achieve this, we rewrite the full Hamiltonian in the second-quantized form, using the basis of eigenstates of $H_{\rm 2DEG}$ without SO interaction: $\Psi_{n,l,\sigma}(r,\theta)=e^{il\theta}R_{n,l}(r)$, where $n\in \mathbb{N}$ is the radial subband number, $l\in\mathbb{Z}$ is the angular momentum quantum number, and $\sigma=\left\{\uparrow,\downarrow\right\}$ represents the spin. $R_{n,l}(r)$ is the normalized radial part of the wavefunction, proportional to the Laguerre polynomials (see Appendix~\ref{SM:Model} for full expressions).

Within this basis, the SO interaction only couples a state  $(n,l,\sigma)$  to $(n,l\pm\sigma,\mp\sigma)$. Consequently, the subspace that couples to the lowest-energy state $l=0$ is spanned by the states $\Psi_{\rm l.e.}=(\Psi_{0,0,\uparrow},\Psi_{0,0,\downarrow},\Psi_{0,-1,\uparrow},\Psi_{0,1,\downarrow})$ [see sketch in Fig.~\ref{fig1}(b)]. The Hamiltonian for the 2DEG, restricted to this minimal subspace, becomes
\begin{eqnarray}
	H_{\rm 2DEG} &=& \sum_{l,\sigma\in \mathrm{l.e.}} \epsilon_{0,l,\sigma} c_{l,\sigma}^{\dagger}c_{l,\sigma} \nonumber \\
	&+& t_{\rm SO} \left(c_{0,0,\uparrow}^\dagger c_{0,1,\downarrow}-c_{0,0,\downarrow}^\dagger c_{0,-1,\uparrow}\right)+\mathrm{h.c.},
	\label{Eq:H2DEG_minimal}
\end{eqnarray}
where the sum is restricted to the states $\left\{(0,\uparrow),(0,\downarrow),(-1,\uparrow),(1,\downarrow)\right\}$. Here $c_{n,l,\sigma}^{\dagger}$ creates an electron in a state $\Psi_{n,l,\sigma}$. The parameters are given by $\epsilon_{0,0}=-\mu_0$, $\epsilon_{0,1}=\epsilon_{0,-1}=-\mu_0+\frac{\hbar^2}{m^*l_\omega^2}$, and $t_{\rm SO}=\frac{\alpha_z}{2l_\omega}$, with $\mu_0$ being a tunable chemical potential. We have assumed that the dominant SO coupling term is the one along the strongest confined direction, $\alpha_z$, with the other components negligibly smaller, $\alpha_r=\alpha_\theta\simeq0$.

It is worth noting that this minimal Hamiltonian remains separable by total angular momentum. However, introducing the SC term breaks this symmetry. Specifically, the SC self-energy in the same basis is given by 
\begin{eqnarray}
	&&\Sigma^{(k)}_{\rm SC}(\omega)=  -\sum_{n,l,m,p}\frac{i\Gamma_{n,m,l,p}}{\sqrt{\Delta^2-\omega^2}}\nonumber \\
	\cdot && \left\{\omega\, \mathrm{sinc}\left((p-l)\delta\theta\right)e^{-i(p-l)\theta_k}c^{\dagger}_{n,l,\sigma}c_{m,p,\sigma} \right. \nonumber \\ 
	+ &&\left.\Delta\, \mathrm{sinc}\left((p+l)\delta\theta\right)e^{-i\left(\phi_k+(p+l)\theta_k\right)}c_{n,l,\sigma}c_{m,p,-\sigma}\right\}+\mathrm{h.c.}, \; \; \; \; \;
	\label{Eq:self-energy_general}
\end{eqnarray}
where
\begin{equation}
	\Gamma_{n,m,l,p}\equiv \Gamma_{\rm eff}\int_{l_{\rm J}}^{\infty} dr\; r R_{n,l}(r)R_{m,p}(r),
\end{equation}
with $l_{\rm J}$ the radius of the junction region and $\Gamma_{\rm eff}\equiv\Gamma_{\rm N}\delta\theta$, which we take to be the same for every SC probe. The parameter $\theta_k$ is the mean angle at which the $k$-th SC lead is positioned and $\delta\theta$ the width of its arc angle [see Fig.~\ref{fig2}(a) for a sketch].

This self-energy term couples subbands with different angular momenta because the SC does not extend over the entire circle. However, superconducting pairing correlations are modulated by the $\mathrm{sinc}$ function, which suppresses pairings between subbands with significantly different angular momenta. In particular, if the SC occupies a quarter of the circle, such that $\delta\theta=\frac{\pi}{2}$, intra-subband pairing (i.e., $l=p$) is entirely suppressed except when~$l=p=0$, while pairings between subbands with opposite angular momenta ($l=-p$) are maximized. Similarly, $\Gamma_{n,m,l,p}$ suppresses the coupling for subbands with large differences in radial quantum numbers $n$ and $m$. In fact, for $l_{\rm J}\rightarrow 0$, this term is only non-zero for $n=m$. 

This justifies that our lowest-energy basis, including now the charge-conjugate partners $\Psi_{\rm l.e.}^{\dagger}=(-\Psi_{0,0,\downarrow}^*,\Psi_{0,0,\uparrow}^*,-\Psi_{0,1,\downarrow}^*,\Psi_{0,-1,\uparrow}^*)$, describes properly the lowest-energy physics of the system. Assuming that $\delta\theta=\frac{\pi}{2}$, and taking the limit $\omega\rightarrow 0$ to simplify the discussion, we obtain
\begin{eqnarray}
	\Sigma^{(k)}_{\rm SC}(\omega\rightarrow0)\rightarrow e^{-i\phi_k}\left[\Gamma_{0,0}\left( c_{0,0,\uparrow}c_{0,0,\downarrow}+c_{0,0,\downarrow}c_{0,0,\uparrow}\right)\right. \nonumber \\
	\left. +\Gamma_{1,1}\left( c_{0,-1,\uparrow}c_{0,1,\downarrow}+c_{0,1,\downarrow}c_{0,-1,\uparrow}\right)\right.\nonumber \\
	\left. +\Gamma_{0,1}\left( e^{-i\theta_k}c_{0,0,\uparrow}c_{0,1,\downarrow}+e^{i\theta_k}c_{0,0,\downarrow}c_{0,-1,\uparrow}\right)\right]+\mathrm{h.c.}, \ \
\end{eqnarray}
with $\Gamma_{0,0}=\Gamma_{\rm N}$, $\Gamma_{0,1}=\Gamma_{\rm N}\frac{2}{\pi}\left(\frac{l_{\rm J}}{l_{\omega}}\right)$, and $\Gamma_{1,1}=\Gamma_{\rm N}\left(\frac{l_{\rm J}}{l_{\omega}}\right)^2$. The terms proportional to $\Gamma_{0,0}$ and $\Gamma_{1,1}$ describes conventional pairings between subbands with opposite quantum numbers. But the third term introduces superconducting correlations between subbands with mismatched angular momenta, $l=0$ and $l=\pm1$. This term, together with the SO interaction, break all the energy-level degeneracies of the system and help driving the system to an  effective $p$-wave superconducting phase. Notably, it depends on both the superconducting phase, $\phi_k$, and the mean position of the SC lead, $\theta_k$. The underlying physics is therefore reminiscent of that of full-shell nanowires~\cite{Vaitiekenas:Science20, Penaranda:PRR20, Paya:PRB24}.

\subsection{Results}
\label{Sec:2B}

We solve the above Hamiltonian [Eqs.~\eqref{Eq:H2DEG_minimal} and ~\eqref{Eq:self-energy_general}] for the lowest-energy set $\left\{(0,\uparrow),(0,\downarrow),(-1,\uparrow),(1,\downarrow)\right\}$ with a finite $\omega$. We use the parameters for a Ge-based heterostructure~\footnote{These parameters for Ge are estimated from Ref.~\onlinecite{Bosco:PRA22} in the absence of an electric field, and are in agreement with experimental observations~\cite{Lodari:PRB19}.}, $m^*=0.08m_0$ (with $m_0$ the free-electron mass) and $\alpha=20$~meV$\cdot$nm, which translates into $\epsilon_{0,1}=0.38$~meV and $t_{\rm SOC}=0.2$~meV. Additionally, we fix $\mu_0=0$ and $\Gamma_{\rm N}=2\Delta=0.4$~meV, and conveniently choose $l_{\rm \omega}=l_{\rm J}=50$~nm; unless otherwise specified. Later we will discuss the behavior of the system when these parameters are changed. 

\begin{figure}
\includegraphics[width=1\columnwidth]{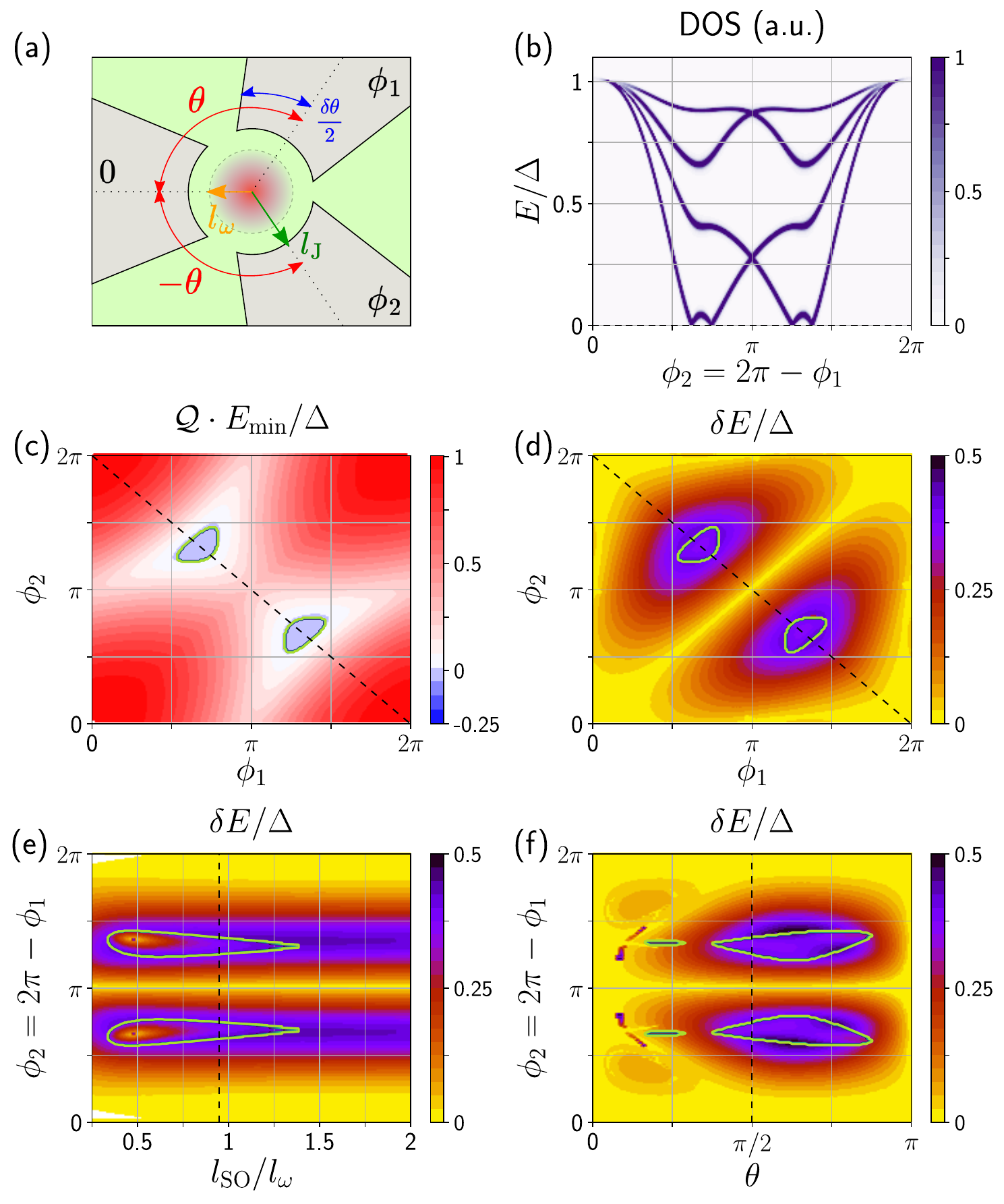}
\caption{\textbf{\label{fig2} Single three-terminal Josephson junction. (a)} Schematics of a 3TJJ. The upper and bottom SCs have a phase $\phi_1$ and $\phi_2$ (we choose the third to be zero), and are positioned at an angle $\theta$ from the middle SC. The harmonic confinement $l_\omega$ and the distance of the 3TJJ center to each SC, $l_{\rm J}$, are not necessarily the same. {\bf (b)} Density of states (DOS) vs the energy $E$ and $\phi_1$, locking $\phi_2=2\pi-\phi_1$. {\bf (c)} Lowest energy level $E_{\rm min}$ (times the parity $\mathcal{Q}$ and normalized to the parent gap $\Delta$) vs $\phi_1$ and $\phi_2$. Zero-energy crossings are marked with a green contour. The dashed line is the phase-path followed in (b). {\bf (d)} Energy difference between the lowest and first-excited energy states, $\delta E$, for the same parameters as in (c). {\bf (e)} $\delta E$ vs the same phases as in (b) and vs the ratio SO length, $l_{\rm SO}=\frac{\hbar^2}{\alpha m^*}$, over harmonic confinement length, $l_{\omega}$. The dashed line correspond to the same cut (and values) than the dashed line in (c, d). {\bf (f)} Same as in (e) but vs the mean angular position $\theta$ of the SCs. The dashed line correspond to the same cut (and values) than the dashed line in (c-e). For all the simulations, we use the following parameters (unless they are varied in the subplot): $m^*=0.08m_0$, $\alpha=20$~meV$\cdot$nm ($l_{\rm SO}\simeq47$~nm), $\mu_0=0$, $\Gamma_{\rm N}=2\Delta=0.4$~meV, $l_{\rm \omega}=l_{\rm J}=50$~nm, $\delta\theta=\frac{\pi}{2}$, and $\theta=\frac{\pi}{2}$.}
\end{figure}

Figure~\ref{fig2}(b) shows the density of states (DOS) as $\phi_1$ varies and $\phi_2$ is set to $\phi_2=2\pi-\phi_1$. We choose $\phi_3=0$ without loss of generality. As expected, the four (positive) energy levels exhibit a splitting, except at the time-reversal-invariant points $\phi_1=\left\{0,\pi\right\}$. Within the range $\phi_1\in\left[0,\pi\right]$, a single energy level crosses zero when the phases align to form a vortex, specifically around $\phi_1=2\pi/3$. For the long-junction $l_{\rm J} \gtrsim l_{\rm SO} $ regime studied here, this level crosses zero energy twice in this range due to the SO interaction. In this regime, the spin and orbital character of each level are locked, so different spins accumulate distinct phase windings similarly to the Aharonov–Casher effect~\cite{Aharonov:PRL84}. Consequently, they disperse differently with the phase and split in energy at finite phase, crossing zero-energy slightly below or above $\phi_1=2\pi/3$, depending on the pseudo-spin. In the complementary phase range, $\phi_1\in\left[\pi,2\pi\right]$, the chirality of the vortex reverses, leading to the formation of an anti-vortex.

To better understand the phase-space extension of these ZEM, we present in Fig.~\ref{fig2}(c) the energy of the lowest-energy mode $E_{\rm min}$ multiplied by the system's parity $\mathcal{Q}(\left\{\phi_i\right\})$, determined through~\cite{Riwar:NatCom16}
\begin{equation}
  \mathcal{Q}(\left\{\phi_i\right\}) = \mathrm{sign}\left\{\mathrm{Pf}\left(\Theta H(\left\{\phi_i\right\})\right)\right\},  
\end{equation}
where $ \mathrm{Pf}(\cdot)$ is the Pfaffian and $\Theta$ the time-reversal operator, written $\Theta=\sigma_y\tau_y\mathcal{K}$ in our basis ($\mathcal{K}$ is the charge-conjugation operator). The dashed line shows the phase-path explored in Fig.~\ref{fig2}(b). The ZEMs, extracted from the spectrum, are outlined with a green contour and agree with the ground-state parity transition. Inside this contour, a single vortex is occupied and the system is in an odd ground state. The energy separation between the two lowest-energy levels is maximum precisely close to the zero-energy crossing, as it can be appreciated in Fig.~\ref{fig2}(d).

To unambiguously detect these ZEMs, it is essential to maximize both the energy splitting, $\delta E$, and the phase-space region over which they occur (i.e., maximize the green contour). To this end, we examine how $\delta E$ and the ZEM position depend on the nanodevice parameters. Figure~\ref{fig2}(e) shows the evolution of $\delta E$ with respect to the ratio $l_{\rm SO}/l_{\omega}$, with $l_{\rm SO}=\frac{\hbar^2}{\alpha m^*}$, along the phase path $\phi_2=2\pi-\phi_1$, which maximizes both quantities. Interestingly, we observe that ZEMs are absent when $l_\omega$ is either above approximately $2 l_{\rm SO}$ or below $\frac{2}{3}l_{\rm SO}$. To obtain ZEMs and maximize $\delta E$, the tunneling region should be confined such that $l_{\rm \omega}\simeq l_{\rm SO}$, allowing the SO energy to match the interband spacing. For a Ge-based heterostructure, since $m^*=0.08m_0$ and $\alpha=20$~meV$\cdot$nm, this implies that $l_{\rm SO}\simeq47$~nm [dashed line in Fig.~\ref{fig2}(e)], and thus one should choose $l_{\omega}\simeq 50$~nm, as we have done for Figs.~\ref{fig2}(b-d). If instead of Ge one uses the parameters of InAs, then the optimal confinement is $l_{\omega}\simeq200$~nm, since the effective mass is four times smaller for InAs than for Ge (assuming the same SO coupling).

In Fig.~\ref{fig2}(f) we perform the same analysis but this time as a function of the mean position of the SCs, $\theta$. Interestingly, we observe an intricate dependence of the vortex formation on the position of the SCs. It appears that, in order to create a vortex, the SCs must be distributed in a relatively even manner, somewhere between the angles $\left[\frac{\pi}{2},\pi\right]$. 

We conduct a similar analysis for the remaining parameters, as detailed in the Appendix~\ref{SM:3TJJ}. Our results indicate that to maximize the splitting, the potential confinement and the 3TJJ size should be of comparable magnitude, i.e., $l_{\rm J}\simeq l_{\rm \omega}$. Additionally, a too strong-coupling to the SCs, $\Gamma_{\rm N}$, leads to a significant metallization of the 2DEG, completely losing the ZEMs.

\section{Double three-terminal Josephson junction}
\label{Sec:2}

\subsection{Model}

We now investigate the behavior of a double 3TJJ. The Hamiltonian consists of two copies of the previous Hamiltonian: one $H_{\rm L}(\phi_1,\phi_2,\phi_3)$ positioned on the left with superconducting contacts placed at $(\theta_1, \theta_2, \theta_3)=(\frac{\pi}{2},-\frac{\pi}{2},\pi)$, and another one at the right $H_{\rm R}(\tilde{\phi}_1,\tilde{\phi}_2,\tilde{\phi}_3)$ with contacts at $(\tilde{\theta}_1, \tilde{\theta}_2, \tilde{\theta}_3)=(\frac{\pi}{2},-\frac{\pi}{2},0)$. These two are coupled to each other through a tunneling Hamiltonian $H_{\rm J}$. The double 3TJJ Hamiltonian then reads
\begin{eqnarray}
	H &=& H_{\rm L}(\phi_1,\phi_2,\phi_3)+ H_{\rm R}(\tilde{\phi}_1,\tilde{\phi}_2,\tilde{\phi}_3) + H_{\rm J}, \\
	H_{\rm L/R} &=& H_{\rm 2DEG,L/R}  + \sum_k^{1,2,3} \Sigma^{(k)}_{\rm SC,L/R}(\omega), \\
	H_{\rm J} &=& - \sum_\sigma \sum_{n,l}\sum_{m,p} J_{n,m,l,p} c^{\dagger}_{n,l,\sigma,\mathrm{L}} c_{m,p,\sigma,\mathrm{R}} + \mathrm{h.c.}, \;\;\;\;\;
	\label{Eq:full_minimal_H}
\end{eqnarray}
where $H_{\rm 2DEG,L/R}$ and $\Sigma^{(k)}_{\rm SC,L/R}(\omega,\phi_1,\phi_2,\phi_3)$ are given by Eqs.~\eqref{Eq:H2DEG_minimal}  and~\eqref{Eq:self-energy_general} acting on the left $c^{\dagger}_{n,l,\sigma,\mathrm{L}}$ or right $c^{\dagger}_{n,l,\sigma,\mathrm{R}}$ 3TJJs states. The coupling parameter is
\begin{equation}
	J_{n,m,l,p} = J_{\rm eff} e^{ip\pi}R_{n,l}(l_{\rm QPC})R_{m,p}(l_{\rm QPC}),
\end{equation}
where $J_{\rm eff}$ is the effective hopping parameter. This term, $J_{n,m,l,p}$, describes the effective tunneling between the left and right 3TJJs. It couples all subbands at both sides, but does not mix spin because we are assuming that the QPC is narrow enough so that there are no spin-splitting effects in the QPC region (see Appendix~\ref{SM:Model} for a more detailed discussion). As we discuss below, including a SO interaction coupling in Eq.~\eqref{Eq:full_minimal_H} extends the range of phases in which PMMs can be found, albeit at the cost of enlarging the lowest-energy subspace (note, however, that this term is unavoidably included in the more realistic simulations presented in Sec.~\ref{Sec:3}). Neglecting this SO coupling, since tunneling is typically suppressed unless the energy levels at both sides of the QPC are resonant, we can still focus on the lowest-energy set for each 3TJJ as in the previous section.

\subsection{Results}
We solve numerically the Hamiltonian of Eq.~\eqref{Eq:full_minimal_H} and search for ZEMs. For each 3TJJ, we fix the optimal parameters for a Ge-based heterostructure found in the previous section [the parameters of Fig.~\ref{fig2}(b)], and analyze the behavior of the system vs $J_{\rm eff}$ and the phases. Since the phase-space is now spanned by five independent phases, we look for ways to reduce this space. 

First, we fix the phases of the left 3TJJ in the same way as we did in the previous section, namely $(\phi_1,\phi_2,\phi_3)=(\phi_1,2\pi-\phi_1,0)$. This choice maximizes the separation between the two lowest-energy levels of the left 3TJJ. For the right 3TJJ, we fix the same configuration of phases, but we add a global phase $\delta\phi$ with respect to the left 3TJJ, this is $(\tilde{\phi}_1,\tilde{\phi}_2,\tilde{\phi}_3)=(\tilde{\phi}_1+\delta\phi,2\pi-\tilde{\phi}_1+\delta\phi,\delta\phi)$. Notice that this setup is somehow similar to the one studied in Ref.~\onlinecite{Samuelson:PRB24}, but breaking TRS on each QD with the aid of the 3TJJ configuration instead of applying a magnetic field or relying on electron-electron interactions.

\begin{figure}
	\includegraphics[width=0.98\columnwidth]{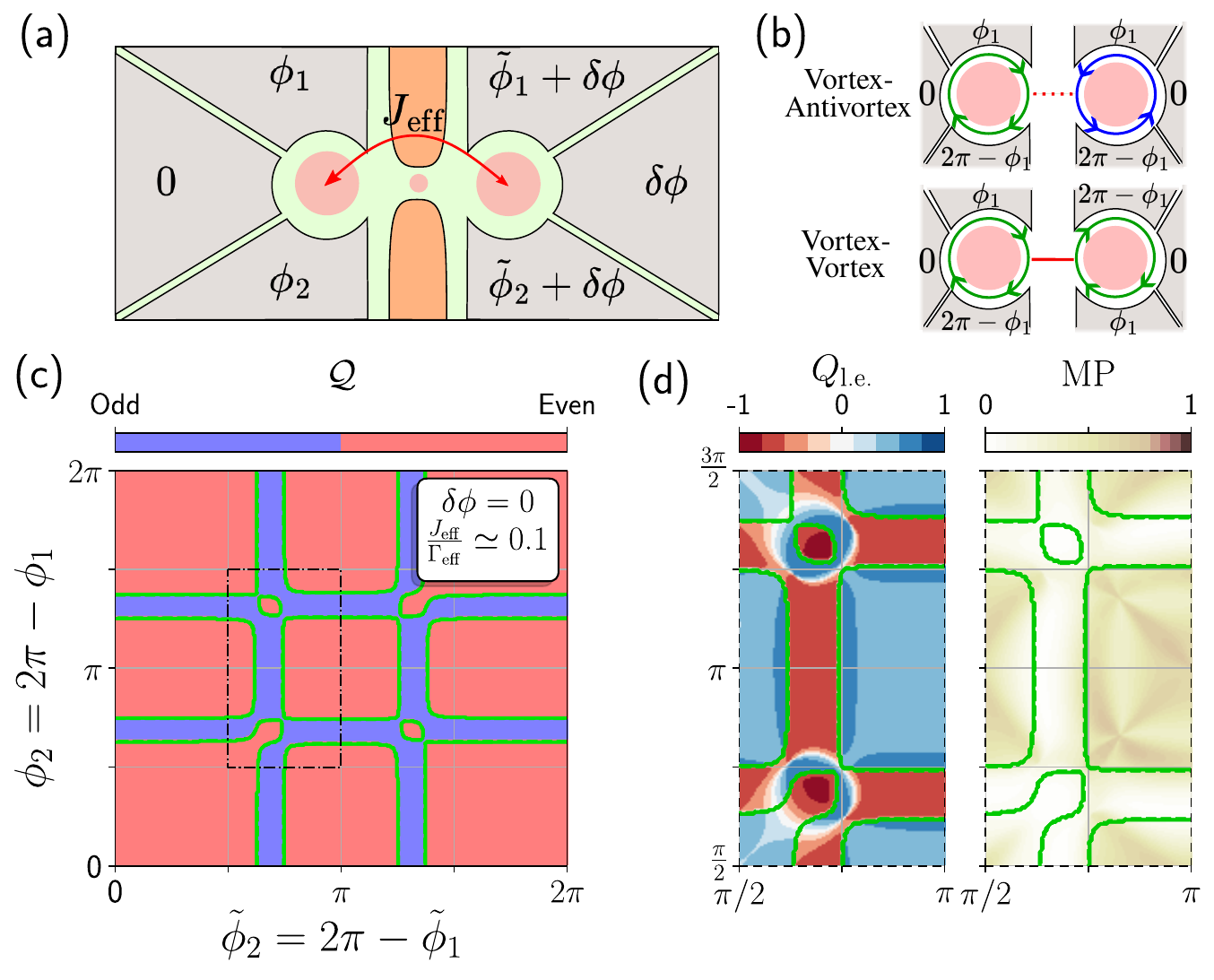}
	\caption{\textbf{\label{fig3} weak-coupling $\mathbf{J_{\rm eff}}$. (a)} Schematics of a double 3TJJ. An effective coupling $J_{\rm eff}$ between both 3TJJs is included in the model. {\bf (b)} Representation of the two possible coupling scenarios: either a vortex and antivortex couple or two vortices do; depending on the phase configuration on each 3TJJ. The vortex-vortex coupling is naturally favored since $J_{\rm eff}$ is spin-conserving. {\bf (c)} Parity of the ground state $\mathcal{Q}$ vs the phases of the left ($\phi_1$ and $\phi_2$) and right ($\tilde{\phi}_1$ and $\tilde{\phi}_2$) 3TJJs, in the weak-coupling limit $J_{\rm eff}\simeq0.1\Gamma_{\rm eff}$. We fix $\delta\phi=0$. The green line depicts the zero-energy crossings. {\bf (d)} Charge of the lowest-energy mode $Q_{\rm l.e.}$ (left) and Majorana polarization MP (right), for the area inside the dashed rectangle of (c). Notice that the MP is always smaller than 0.5.}
\end{figure}

\subsubsection{weak-coupling}

We begin by fixing $\delta\phi=0$ and considering the case of weak-coupling between these two 3TJJs, specifically $J_{\rm eff}\simeq0.1\Gamma_{\rm eff}$. In this regime, the system behaves approximately as two weakly-coupled 3TJJs, which is useful for understanding its phase-space behavior. Figure~\ref{fig3}(c) illustrates the parity of the ground state, $\mathcal{Q}$, as a function of the phase configuration on the left 3TJJ $(\phi_1,2\pi-\phi_1,0)$ and the right 3TJJ $(\tilde{\phi}_1,2\pi-\tilde{\phi}_1,0)$. The green contour depicts the position of the ZEMs which aligns with the odd/even parity transitions. Along one coordinate axis, such as when $\phi_1=0$ and varied $\tilde{\phi}_1$, each 3TJJ individually undergoes vortex or antivortex formation, as described in the previous section, while the other 3TJJ remains unoccupied. It is within the central region, inside the dashed rectangle, that vortex/antivortex formation occurs simultaneously in both 3TJJs, allowing them to hybridize. 

In this weak-coupling regime, even-occupied islands emerge within the odd-occupation contour, corresponding to a doubly occupied vortex state, with one vortex on each 3TJJ. Notably, the two islands at $(\frac{2\pi}{3},\frac{2\pi}{3})$ and at $(\frac{2\pi}{3},2\pi-\frac{2\pi}{3})$ differ in behavior. The former results from the hybridization of a vortex and an antivortex [see Fig.~\ref{fig3}(b)], leading to an anticrossing and the removal of the ZEM. In contrast, the latter arises from the hybridization of two vortices (i.e., with the same vorticity), resulting in a ZEM that extends across both junctions. This is a result of the spin-conserving nature of the coupling between 3TJJs, $J_{n,m,l,p}$, which favors a coupling between states with same pseudo-spin or vorticity. Therefore, the configuration around $(\frac{2\pi}{3},2\pi-\frac{2\pi}{3})$ is the most favorable for obtaining PMM modes. In a more general scenario where a SO coupling is also considered in $J_{n,m,l,p}$, the coupling between a vortex and antivortex provide ZEMs as well. In this scenario, both even-occupied islands appear similar (not shown) and can lead to the formation PMMs.

To gain deeper insight, we compute in the left panel of Fig.~\ref{fig3}(d) the charge of the lowest-energy mode $Q_{\rm l.e.}$ within the dashed rectangle of (c). Notably, certain phase points within this region exhibit ZEMs with zero net charge, especially in specific corners associated with double occupancy. This observation suggests the presence of a ZEM localized at both junction sites without net charge. This is a necessary condition for MBSs but not sufficient.

A way to determine if these ZEMs are indeed well-behaved PMM modes is by calculating the so-called Majorana polarization (MP) as in Refs.~\onlinecite{tsintzis_creating_2022, Samuelson:PRB24} 
\begin{equation}
	\mathrm{MP}_{\rm L/R} = \frac{\left|\sum_{\sigma, s}\left<e\left|\gamma_{\sigma,s,\mathrm{L/R}}\right|o\right>^2\right|}{\sum_{\sigma, s}\left|\left<e\left|\gamma_{\sigma,s,\mathrm{L/R}}\right|o\right>^2\right|},
\end{equation}
where $\left|o\right>$ and $\left|e\right>$ are the odd and even ground states, and $\gamma_{\sigma,+,\mathrm{L/R}}=c_{\sigma,\mathrm{L/R}}+c_{\sigma,\mathrm{L/R}}^\dagger$ and $\gamma_{\sigma,-,\mathrm{L/R}}=i\left(c_{\sigma,\mathrm{L/R}}-c_{\sigma,\mathrm{L/R}}^\dagger\right)$ are the Majorana operators acting on the left (L) or right (R) 3TJJ. The MP serves as a quality factor by quantifying the spatial separation of particle and hole components within a certain mode, a key feature of MBSs. When $\mathrm{MP}\sim 1$, the mode is a perfect mixture of electron and hole, and thus it has a strong Majorana character. Conversely, if the MP deviates significantly from unity, then the state is less likely to exhibit true Majorana behavior, instead reflecting a more conventional fermionic quasiparticle nature. Since our system is not necessarily symmetric, we take $\mathrm{MP}=(\mathrm{MP}_{\rm L}+\mathrm{MP}_{\rm R})/2$, imposing that $\mathrm{MP}_{\rm L/R}$ must be large in both 3TJJs to obtain a good PMM mode.

In the right panel of Fig.~\ref{fig3}(d) we display the MP for the same phase-space region as in the left panel. The MP shows a complex variation across the phase-space, whose interpretation is only meaningful near the ZEMs, i.e., along the green contours. We observe that the MP reaches a maximum of $\sim0.5$ in this regime, indicating a weak Majorana character for these ZEMs.

\subsubsection{Optimal regime}

\begin{figure}
	\begin{centering}
		\includegraphics[width=0.98\columnwidth]{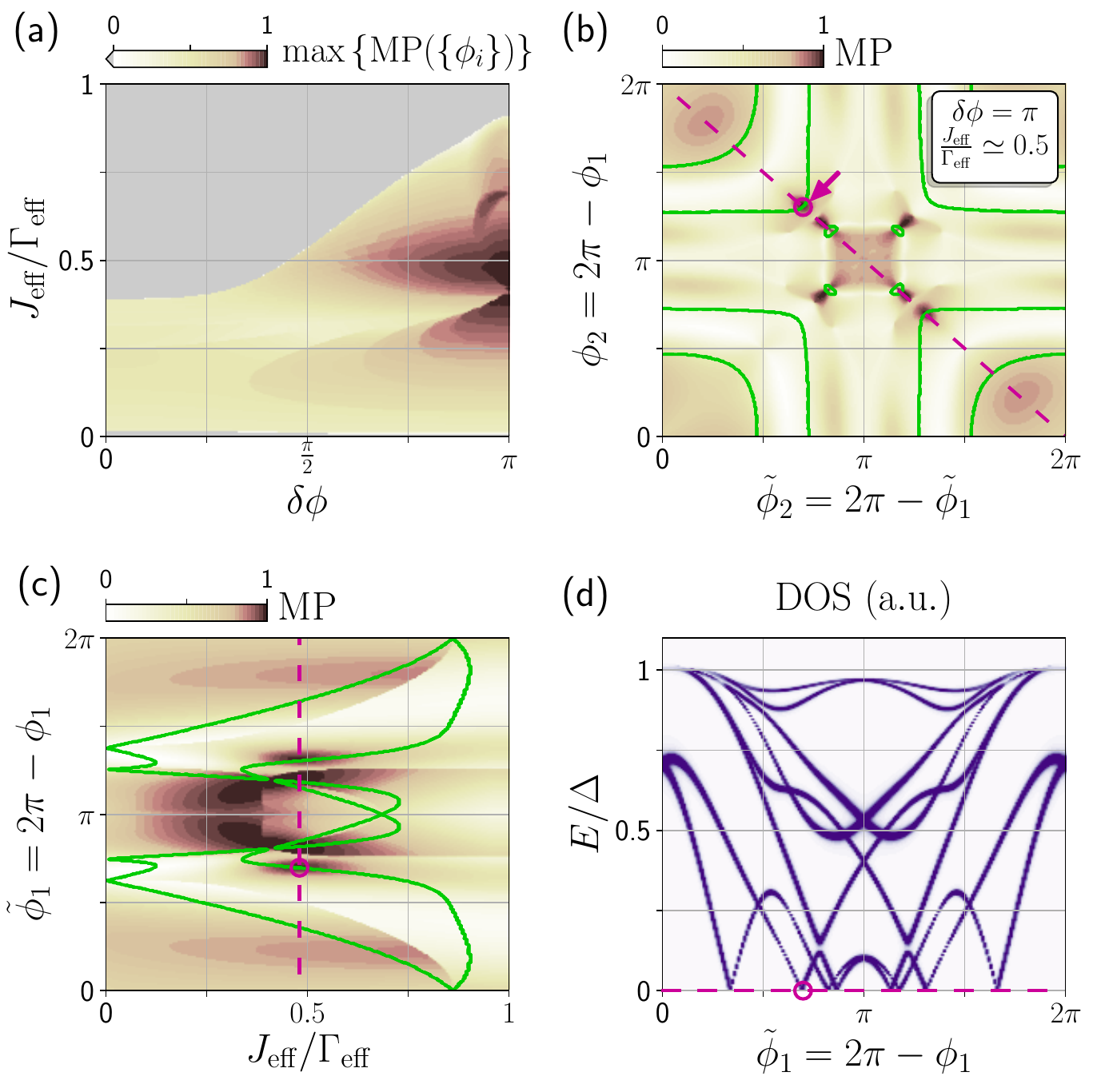}
		\par\end{centering}
	\caption{\textbf{\label{fig4} Optimal parameters. (a)} Maximum of the Majorana polarization (MP) among all the phases $\left\{\phi_i\right\}=\left\{\phi_1,\phi_2,\tilde{\phi}_1,\tilde{\phi}_2\right\}$, fixing different $\delta\phi$ and $J_{\rm eff}/\Gamma_{\rm eff}$, with $\Gamma_{\rm eff}=\Gamma_{\rm N}\delta\theta$. In the gray shaded area no ZEMs are obtained. {\bf (b)} MP vs the phases in both 3TJJs, fixing the optimal parameters that gives $\mathrm{MP}\sim 1$ in (a), i.e., $\delta\phi=\pi$ and $J_{\rm eff}\simeq\frac{\Gamma_{\rm eff}}{2}$.  {\bf (c)} MP vs $J_{\rm eff}/\Gamma_{\rm eff}$ and the phases $(\phi_1,\tilde{\phi}_1)=(\phi_1,2\pi-\phi_1)$ [along the purple dashed line in (b)]. The complementary phases are set to $(\phi_2,\tilde{\phi}_2)=(2\pi-\phi_1,\phi_1)$, and $\delta\phi=\pi$. {\bf (d)} DOS vs the energy $E$ (normalized to the parent gap $\Delta$) along the phase path shown in (b, c) with a dashed line. The purple circle marks the ZEM that we identify as a PMM mode.}
\end{figure}

To optimize the conditions for PMM modes, we analyze the maximum MP of all the ZEMs across the whole phase space $\left\{\phi_i\right\}$ while varying the relative phase $\delta\phi$ and the coupling strength $J_{\rm eff}$. The results are summarized in Fig.~\ref{fig4}(a). The gray-shaded area correspond to the parameter space where no ZEM is observed at all. We identify the optimal parameters to be $\delta\phi=\pi$ and $J_{\rm eff}\simeq\frac{\Gamma_{\rm eff}}{2}$, which provides a nearly perfect MP of $\sim 0.99$. This indicates a strong Majorana character for the ZEMs at these settings, and remarkably, seems to be pretty robust against phase or coupling variations. Notice that the condition $J_{\rm eff}\simeq\frac{\Gamma_{\rm eff}}{2}$ resembles somehow the condition $t=\Delta$ to obtain PMM modes in the Kitaev chain, being $t$ the hopping between sites and $\Delta$ the inter-site pairing in that model. On the other hand, $\delta\phi=\pi$ ensures a strong phase difference an a perfect alignment of the vortices at both junctions.

To know at which exact phase configuration $\left\{\phi_i\right\}$ the maximum MP happens, we present in  Fig.~\ref{fig4}(b) the MP as a function of the phases at these optimal parameters, $\delta\phi=\pi$ and $J_{\rm eff}\simeq\frac{\Gamma_{\rm eff}}{2}$ (we refer to Appendix~\ref{SM:double-3TJJ} to see this phase diagram for different parameters). We observe that the maximum MP (in a ZEM) occurs at some point along the phase-path marked with an purple dashed line, i.e., the phase configuration $(\phi_1,\tilde{\phi}_1)=(\phi_1,2\pi-\phi_1)$ with the complementary phases $(\phi_2,\tilde{\phi}_2)=(2\pi-\phi_1,\phi_1)$. This choice of phases actually ensures that the left and right 3TJJs have matching vorticities, see Fig.~\ref{fig3}(b) (both are vortices and not vortex and antivortex). The ZEM with $\mathrm{MP}\sim 1$, i.e., the PMM mode, is actually at the phase configuration $(\phi_1,\tilde{\phi}_1)=(\frac{3\pi}{4},2\pi-\frac{3\pi}{4})$, denoted with an purple circle in (b) (and pointed with an arrow).

To experimentally detect these PMM modes, the best strategy is thus to set the relative phase $\delta\phi=\pi$ together with this phase-path, $(\phi_1,\tilde{\phi}_1)=(\phi_1,2\pi-\phi_1)$, and tune $J_{\rm eff}$ via the QPC gate potential. By monitoring the local and non-local conductances, one can determine the presence of MBSs, as we detail in the next section. Here, it is more straightforward to analyze the MP directly. In Fig.~\ref{fig4}(c), we plot both the ZEMs (marked by green contours) and the MP (colorbar) as functions of $J_{\rm eff}$ and the phases $(\phi_1,\tilde{\phi}_1)=(\phi_1,2\pi-\phi_1)$ [the purple dashed line in (b)]. As seen in the figure, at weak-coupling, the system behaves as discussed in Fig.~\ref{fig3}. However, as the coupling $J_{\rm eff}$ increases, the ZEMs shift in phase space in a complex manner, with some values achieving a near-perfect Majorana polarization of $\mathrm{MP}\simeq 1$. At these points, the non-local conductance should theoretically vanish, while the local conductance exhibits a pronounced peak, providing an experimental signature of MBSs.

In Fig.~\ref{fig4}(d), we show the DOS along the same phase path, the one marked with an purple dashed line in panel (b), as a function of energy. The second ZEM, marked with an purple circle, corresponds to a nearly perfect PMM mode. A clear minigap of $\sim 0.25\Delta$ is observed at this point, which is sufficiently large to be experimentally detected. 

\section{Full-Hamiltonian simulations}
\label{Sec:3}
Having established the optimal conditions for achieving PMM modes for the simplified model system, we proceed to examine a more realistic configuration, focusing on both the robustness of the PMM modes and their experimental detectability. We consider the continuum Hamiltonian described by Eq.~\eqref{Eq:H_continuum2} applied to a more general junction. To simulate this system, we discretize the 2D Hamiltonian on a rectangular grid and solve it using finite difference methods, using the the routines implemented in Refs.~\onlinecite{Escribano:software, Escribano:tesis}. Unlike the idealized circular tunneling region, here we assume a rectangular junction geometry and incorporate disorder, which is modeled as a random onsite chemical potential within the 2DEG, in order to slightly break the left/right symmetry. Note that the effect of this asymmetry (or any other arising from the geometrical configuration of the nanodevice) merely renormalizes the chemical potential at each junction and their coupling to the superconductors. This results in a shift of the parameter values at which ZEMs and PMMs can be found, which must be explored regardless in both, our simulations and in experiments.

\begin{figure}
	\begin{centering}
		\includegraphics[width=0.98\columnwidth]{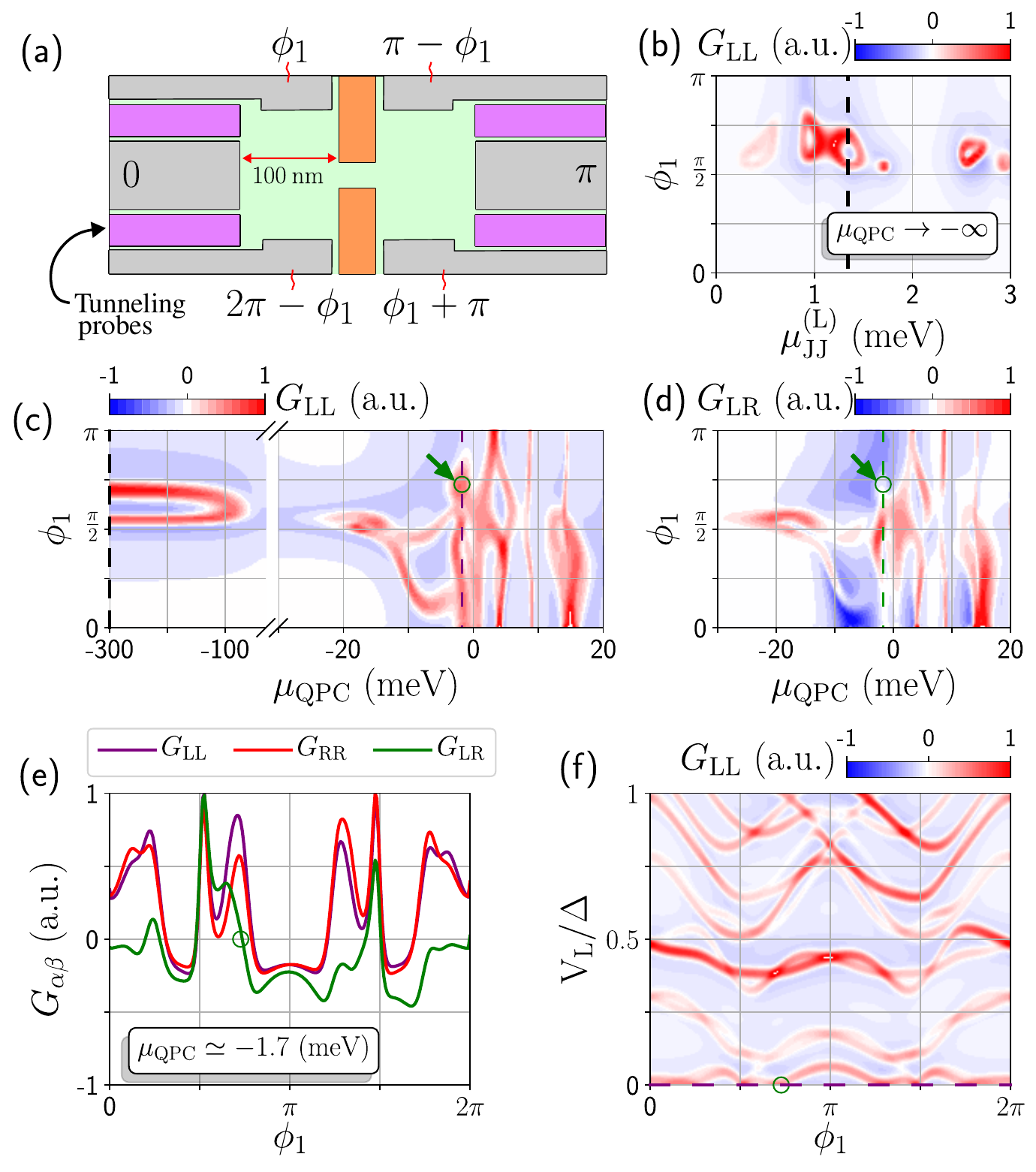}
		\par\end{centering}
	\caption{\textbf{\label{fig5} Full-Hamiltonian simulations. (a)} Schematic of the realistic device used in the continuum model simulations. We include normal probes (in purple) to perform transport simulations. We set the phases as illustrated, which maximizes the odds of finding PMM modes. {\bf (b)} Local differential conductance on the left, $G_{\rm LL}\equiv\frac{dI_{\rm L}}{dV_{\rm L}}$, in arbitrary units, as a function of the chemical potential in the left 3TJJ $\mu_{\rm J}$ and $\phi_1$. Here, the QPC gate voltage is set to a large negative value ($\mu_{\rm QPC}=-300$~meV), effectively decoupling the left and right 3TJJs. {\bf (c)} Local differential conductance on the left 3TJJ, $G_{\rm LL}$, as a function of $\phi_1$ and the chemical potential in the QPC region, $\mu_{\rm QPC}$. {\bf (d)} Non-local differential conductance, $G_{\rm LR}$, within a zoomed-in region of (c). {\bf (e)} Local and non-local differential conductances for the $\mu_{\rm QPC}$ value marked with the colored dashed lines in (d, c). {\bf (f)} Local differential conductance vs $\phi_1$ for the same $\mu_{\rm QPC}$ value as in (e) vs the energy of the left probe $V_{\rm L}$. The green circle marks the zero-energy state that corresponds to a PMM mode. Temperature is fixed to $T=100$~mK in these simulations.}
\end{figure}

Our configuration, illustrated schematically in Fig.~\ref{fig5}(a), includes normal leads (depicted in purple) to perform conductance simulations, which will serve us as a smoking gun for the detection of the PMM modes. The differential conductance is
\begin{equation}
    G_{\alpha\beta} \equiv \frac{dI_{\alpha}}{dV_{\beta}},\quad \forall \, \alpha,\beta=\left\{\mathrm{L,R}\right\},
\end{equation}
where $I_{\alpha}$ is the current through the $\alpha$ (left or right) leads, and $V_{\beta}$ the voltage at the $\beta$ (left or right) ones. The local conductance is when measuring the current through the same lead whose voltage is changed, i.e., $\alpha=\beta$. While the non-local conductance is when measuring the current through the opposite lead where the voltage is changed, $\alpha\neq\beta$. See Appendix~\ref{SM:numerics} for details on simulating this conductance. 

The procedure to create and detect PMM modes, derived from our explanations in the previous section, involves the following steps:

(i) Start by closing the QPC to isolate the left and right 3TJJs. Independently tune the gate potential for each 3TJJ to bring one subband in each tunneling region close to zero energy, setting effectively $\epsilon_{0,0} = 0$. To identify this alignment, measure the local conductance at $V_{\alpha}=0$~V as a function of the gate potential and the phase configuration, specifically setting $(\phi_1, \phi_2, \phi_3) = (\phi_1, 2\pi - \phi_1, 0)$. Conductance peaks will appear when the chemical potential $\mu_{\rm J}$ aligns with a subband energy, as illustrated by the simulations in Fig.~\ref{fig5}(b) only for the left 3TJJ. We observe multiple circular patterns in the parameter space, each associated with a distinct subband. These are circles rather than points due to the spin-splitting effects described in Fig.~\ref{fig2}. A larger circle implies a stronger spin-splitting effect, whereas the absence of circles would suggest that the junction size is not comparable (either too small or to large) to the SO length of the heterostructure. 

(ii) After identifying the best gate potential from the previous step (preferably the one with the maximum spin-splitting), set $\delta \phi = \pi$ and gradually increase the QPC voltage to enhance effectively the coupling $J_{\rm eff}$. To detect PMM modes, measure both local and non-local conductances at $0$~V as functions of the gate potential and phase $\phi_1$, locking the rest to $(\phi_1,\phi_2,\phi_3)=(\phi_1,2\pi-\phi_1,0)$ and $(\tilde{\phi}_1,\tilde{\phi}_2,\tilde{\phi}_3)=(2\pi-\phi_1+\delta\phi,\phi_1+\delta\phi,\delta\phi)$, see the resulting phase configuration in Fig.~\ref{fig5}(a). This specific phase configuration aligns the vortices optimally, creating conditions favorable for PMM mode formation, as explained before. Experimentally, a PMM is identified by a peak in the local conductance at both junctions, together with near-zero non-local conductance. 

To illustrate this, we show in Fig.~\ref{fig5}(c,d) a simulated example on this setup. We show the (c) local and (d) non-local conductances as a function of $\phi_1$ and the chemical potential within the QPC, $\mu_{\rm QPC}$, for the fixed chemical potential in the 3TJJ indicated with a dashed line in Fig.~\ref{fig5}(b). At large negative $\mu_{\rm QPC}$, the QPC is effectively closed, leaving the two 3TJJ decoupled [the situation in (1)]. Conversely, at $\mu_{\rm QPC}\gg\mu_{\rm J}\simeq1.2$~meV the QPC defines a quantum well, creating localized states in the QPC barely sensitive to the phase $\phi_1$. The relevant regime is thus $\mu_{\rm QPC}\lesssim0$, right part on Fig.~\ref{fig5} (c). Figure~\ref{fig5} (d) is already a zoom in this area. Notice that the behavior of the conductance peaks resemble somehow to the ZEM behavior in our minimal model, i.e., the green contour in Fig.~\ref{fig4}(b). With a dotted line, we indicate the phase-path along which we observe a local conductance peak concurring with a vanishing non-local conductance. The green dot (pointed with an arrow) marks exactly the phase at which this occurs, pretty close to the value $\phi_1=\frac{3\pi}{4}$ obtained in the previous section. 

To better illustrate this, we show a cut of the differential conductances in Fig.~\ref{fig5}(e). With the green dot we show the value at which we obtain a zero non-local conductance together with a peak of the local conductance at both sides, which we identify as a PMM mode. The conductances $G_{\rm LL}$ and $G_{\rm LR}$ are not symmetric with respect to $\phi_1=\pi$, as it was in our minimal model, because disorder breaks the left/right symmetry of our nanodevice.

(iii) To further check the protection of these modes, one can measure the local conductance vs the voltage of the probe, which should be proportional to the density of states (and the spectrum). We show this simulation in Fig.~\ref{fig5}(f), finding that the minigap for the identified PMM (green circle) is around $0.08\Delta$. 

We plot the probability density of the optimal PMM mode along the junction as extracted from the exact diagonalization of the full Hamiltonian, which we show in the Appendix~\ref{SM:Continuum}. We observe that the state corresponding to the green circle is perfectly localized at opposite ends of the double-junction, as expected from a PMM mode. Other ZEM at different phases do not exhibit this distinct feature, instead showing a extended distribution at either one or both junctions. In addition, we compute the MP for these eigenstates, finding $\mathrm{MP}\simeq 0.95$ and confirming that they are PMM modes.

\section{Conclusions}

In this work, we have argued that it is possible to achieve poor's man Majorana~(PMM) modes in a double three-terminal Josephson junction~(3TJJ) configuration [see Fig.~\ref{fig1}(a)]. This system does not require the application of any magnetic field but instead a fine-tuning of the superconducting phases of each individual superconductor~(SC).

We derive a minimal model, whose parameters are intimately related with the physical and geometrical parameters of the system, through which we explain the conditions for the formation of a PMM. We identify the optimal device geometry and the phases for which this mode should appear. To this end, we use the Majorana polarization~(MP) as a quality factor, achieving approximately $1$ in the optimal case. Particularly, we find that the size of the junction should be of the order of the spin-orbit~(SO) length that characterizes the underlying 2DEG; the SCs must be distributed somehow evenly along the junction; and the coupling between both 3TJJs, $J_{\rm eff}$, must be approximately the half of the coupling to each SC $\Gamma_{\rm eff}$, i.e., $J_{\rm eff}\simeq\frac{\Gamma_{\rm eff}}{2}$. Under these conditions, we find PMM modes when setting the phases near to $(\phi_1,\phi_2,\phi_3;\tilde{\phi}_1,\tilde{\phi}_2,\tilde{\phi}_3)=(\frac{3\pi}{4},\frac{5\pi}{4},0; \frac{7\pi}{4},\frac{\pi}{4},\pi)$.

From this minimal model, we develop a method to experimentally find and detect PMM modes by relying on local and non-local measurements of the conductance. We test this procedure by computing the spectrum and the conductance for the full (quasi-2D) Hamiltonian in a more general configuration, including disorder that breaks the left/right symmetry of the nanodevice. Following the procedure that we described, we are able to find PMM by looking into the conductance features. We compare them with the numerical diagonalization of the Hamiltonian to verify that the zero-energy modes are certainly PMM modes, finding a MP as good as $0.95$. We find the PMM mode occurring at a phase configuration similar to the one obtained with the minimal model.

In this configuration, we find a minigap of $\sim 0.08\Delta$. Although the minigap is relatively small, we expect the energy levels to exhibit a better resolution because no magnetic field needs to be applied, and thus there are no detrimental effects coming from it. In addition, since these states appear in a strong-coupling regime with the SCs, we expect charging effects to be minimal. Moreover, we believe our double 3TJJ system is scalable to an arbitrary number of junctions, scaling up the number of Kitaev chain sites, as explained in Appendix~\ref{SM:Scalability}. In such a case, the perfect geometrical symmetry imposed on the system analyzed in this work would no longer hold. Nevertheless, we still expect a PMM mode to emerge, albeit at a different value of $\delta\phi$, which would need to be determined through future studies. 

This nanodevice also offers the advantage of working across a broad range of materials. In this manuscript, we use parameters specific to Ge-based heterostructures for the 2DEG, as these do not require a strong $g$-factor, relying instead on materials with significant SO coupling. Similarly, III-V compound SMs, which leverage the properties of the conduction band, could also be suitable, although they would require larger junction sizes due to their relatively smaller Rashba-type SO coupling and/or smaller effective masses (see discussion in Sec.~\ref{Sec:2B}). In addition, a recent study~\cite{vezzosi:arxiv2024} has proposed heterostructures that exploit the valence band of III-V SMs, like GaSb-InP heterostructures, achieving SO couplings comparable to Ge-based heterostructures but potential better mobilities. For the SCs, the choice of materials is equally versatile, as long as it is a conventional $s$-wave superconductor capable of inducing a hard gap in the SM, such as Al~\cite{Krogstrup:NM15, Delaforce:AdvanceMaterials21}, Pb~\cite{Kanne:NatNano21}, Sn~\cite{Goswami:ACSnano23, Goswami:NanoLet23}, Nb~\cite{Hato:IOP92, Gusken:Nanoscale17, Langa:IAP24} or Pt~\cite{Tosato:ComMat23}.

\acknowledgments{We acknowledge useful discussions with Magnus R. Lykkegaard, Nadav Drechsler, and Omri Lesser. This research was funded in part by the European Research Council (Grant No.~856526), by the DFG Collaborative Research Center (CRC) 183 Project No.~277101999, and by ISF Grants No.~1914/24 and No.~2478/24.

\newpage
\onecolumngrid
\appendix

\newpage

\section{Device Scalability}
\label{SM:Scalability}

Our proposed nanodevice can be scaled into a longer chain comprising an arbitrary number of three-terminal Josephson junctions (3TJJs). Analogously to the Kitaev chain, extending the length of the chain will enhance the topological protection and improve the robustness against perturbations~\cite{Kitaev:PU01, Miles:PRB24}. Nonetheless, Ref.~\onlinecite{Luethi:arxiv24} has shown that is not guaranteed that all PMM modes are connected to topologically-protected MBSs in longer chains. Hence, the optimal number of 3TJJ that maximizes the topological minigap must be found.

An example configuration is illustrated in Fig.~\ref{fig6}. In this geometry, the mean-angular positions of the superconducting probes (gray) are fixed at $\theta=\frac{2\pi}{3}$, which actually corresponds to the optimal configuration, while the angular width is kept at $\delta\theta\simeq\frac{\pi}{2}$. The quantum point contacts (QPCs, orange), which separate individual 3TJJs, are positioned at a different symmetry-points relative to the superconductors. As a result, the previously optimal superconducting phase difference $\delta\phi=\pi$ may no longer apply but is likely to shift to a value close to $\pi$. Notice this shift must be cumulative along the 3TJJs, i.e., one should fix $\delta\phi$ in the second 3TJJ, $2\delta\phi$ in the third, and so on. Notably, this configuration also enables the integration of local normal probes (purple) within each 3TJJ, facilitating comprehensive measurements of both local and non-local conductances along the chain.

\begin{figure}
	\begin{centering}
		\includegraphics[width=0.7\columnwidth]{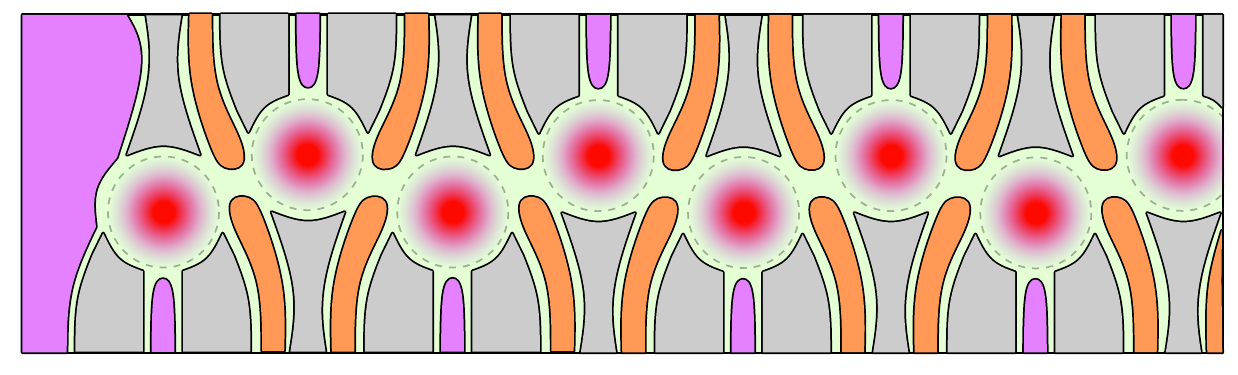}
		\par\end{centering}
	\caption{\textbf{\label{fig6} System's scalability.} Sketch illustrating one possible way to scale our nanodevice into a longer Kitaev chain (a semi-infinite chain is depicted). The underlying semiconductor layer, where the 2DEG is confined, is shown in green. Superconducting probes are depicted in gray, fixed at $\theta=\frac{2\pi}{3}$ with a width $\delta\theta\simeq\frac{\pi}{2}$ within each three-terminal Josephson junction (3TJJ). Normal gates, used to define the quantum point contacts (QPCs) that regulate coupling between the 3TJJs, are shown in orange. Purple probes correspond to the normal probes that enable measurements of local and non-local conductances at various points along the chain. }
\end{figure}

\section{Derivation of the Minimal Model}
\label{SM:Model}
\subsection{Minimal model for a single multi-terminal Josephson junction}
We begin by describing a single multi-terminal Josephson junction. The starting point is a continuum Hamiltonian that models the 3D planar heterostructure
\begin{equation}
    H_{\rm full}(\vec{r}) =\left(\vec{k}\frac{\hbar^2}{2m^*(\vec{r})}\vec{k}-V(\vec{r})\right)\sigma_0\tau_z+\frac{1}{2}\left[\vec{\alpha}(\vec{r})\cdot(\vec{\sigma}\times\vec{k})+(\vec{\sigma}\times\vec{k})\cdot\vec{\alpha}(\vec{r})\right]\tau_z + \Delta(\vec{r})\left[\cos\left(\phi(\vec{r})\right)\tau_x+\sin\left(\phi(\vec{r})\right)\tau_y\right]\sigma_0,
\end{equation}
where $\sigma_i$ and $\tau_i$ are the Pauli matrices in spin and Nambu space, respectively. The position and momentum operators are represented by $\vec{r}=(x,y,z)$ and $\vec{k}=(k_x,k_y,k_z)$. The parameter $m^*(\vec{r})$ is the effective mass, $V(\vec{r})$ the electrostatic potential, $\alpha(\vec{r})$ the spin-orbit (SO) coupling, $\Delta(\vec{r})$ the superconducting pairing correlations, and $\phi(\vec{r})$ the superconducting phase. These parameters are spatial dependent as they take different values depending on whether $\vec{r}$ lies in the semiconductor~(SM) or the superconductor~(SC).

To simplify the problem, we integrate out the degrees of freedom associated with the SC regions, incorporating the effects of superconductivity into the SM region as a self-energy term. Furthermore, we assume that the active layer of the SM is sufficiently thin such that only a single transverse subband is occupied. Under these assumptions, the position vector $\vec{r}$ now applies exclusively to the SM and is restricted to the planar coordinates $(x,y)$. The resulting Hamiltonian reads
\begin{eqnarray}
    H_{\rm full}(\vec{r},\omega) & = & H_{\rm 2DEG}(\vec{r}) + \sum_k \Sigma_{\rm SC}^{(k)}(\vec{r},\omega),
    \label{Eq:H_total}
\end{eqnarray}
where $H_{\rm 2DEG}$ describes the two-dimensional electron/hole gas~(2DEG/2DHG) in the SM, while $\Sigma_{\rm SC}^{(k)}$ represents the self-energy contribution from every $k$-th SC probe. These terms are defined as:
\begin{gather}
    H_{\rm 2DEG}(\vec{r}) = H_0(\vec{r}) + H_{\rm SO}, \\
    H_0(\vec{r}) = \left(\frac{\hbar^2\vec{k}^2}{2m^*}-V(\vec{r})\right)\tau_z, \\
    H_{\rm SO} = \frac{1}{2}\left[\vec{\alpha}\cdot(\vec{\sigma}\times\vec{k})+(\vec{\sigma}\times\vec{k})\cdot\vec{\alpha}\right]\tau_z,  \\
    \Sigma_{\rm SC}^{(k)}(\vec{r},\omega) = -\frac{i\Gamma_{\rm N}(\vec{r})}{\sqrt{\Delta^2-\omega^2}}\left(\omega\tau_0 +\Delta\cos(\phi_k)\tau_x-\Delta\sin(\phi_k)\tau_y\right)\sigma_0. \label{Eq:Sigma_app}
\end{gather}
The tunneling ratio $\Gamma_{\rm N}(\vec{r})$ takes a constant value (fixed by the material and interface properties) beneath each SC region and vanishes elsewhere. We assume here that each $k$-th SC is characterized by a constant superconducting phase $\phi_k$.

To develop a lowest-energy model, we rewrite the Hamiltonian in the basis spanned by the eigenfunctions of $H_0$. For this purpose, we assume a harmonic potential confinement within the junction, i.e., $V(\vec{r})=\frac{1}{2}m^*\omega_{\rm o}^2r^2$ (with $\omega_{\rm o}$ the oscillator frequency), and express the Hamiltonian in polar coordinates $\vec{r}=(r,\theta)$,
\begin{equation}
    H_0(\vec{r})=-\frac{\hbar^2}{2m^*}\left(\partial_r^2+\frac{1}{r}\partial_r+\frac{1}{r^2}\partial_\theta^2\right)\tau_z+\frac{1}{2}m^*\omega_{\rm o}^2r^2\tau_z.
\end{equation}
The solutions to this Hamiltonian, energy and eigenfunctions, with $l_\omega=\sqrt{\frac{\hbar}{m^*\omega_{\rm o}}}$, are given by
\begin{eqnarray}
    \epsilon_{n,l}=\frac{\hbar^2}{m^*l_\omega^2}(2n+\left|l\right|+1), \quad \Psi_{n,l,\sigma}(r,\theta)=e^{il\theta}R_{n,l}(r),
\end{eqnarray}
with the radial component
\begin{eqnarray}
    R_{n,l}(r)=C_{n,l}\left(\frac{r}{l_{\omega}}\right)^{\left|l\right|}e^{-\frac{r^2}{2l_{\omega}^2}}\mathcal{L}_{n}^{\left| l \right|}\left\{\left(\frac{r}{l_\omega}\right)^2\right\}.
\end{eqnarray}
Here, $C_{n,l}$ is a renormalization constant and $\mathcal{L}_{n}^{\left| l \right|}\left\{x\right\}$ the associate Laguerre polynomials, which explicitly they are
\begin{equation}
    C_{n,l}=\frac{\sqrt{2}}{l_\omega}\sqrt{\frac{n!}{(n+l)!}}, \quad 
    \mathcal{L}_{n}^{\left| l \right|}\left\{x\right\}=\frac{x^{-l}}{n!}\left(\frac{d}{dx}-1\right)^l x^{n+l}.
\end{equation}
Using this basis, the Hamiltonian $H_0$ can be written in diagonal form
\begin{eqnarray}
    H_0=\sum_{n,l,\sigma}\epsilon_{n,l}c_{n,l,\sigma}^\dagger c_{n,l,\sigma}+\mathrm{h.c.},
\end{eqnarray}
with $c^\dagger_{n,l,\sigma}$ and $c_{n,l,\sigma}$ the creation and annihilation operators for an electron in the state with radial quantum number $n$, angular $l$ and spin $\sigma$.

We now proceed to express the SO coupling in the basis of the eigenfunctions of $H_0$. Expanding the SO interaction term, we have
\begin{eqnarray}
    \vec{\alpha}\cdot(\vec{\sigma}\times\vec{k}) = \alpha_r\left(-\sin(\theta)k_z\sigma_x + \cos(\theta)k_z\sigma_y-\sigma_zk_\theta\right) +\alpha_z \left(\cos(\theta)k_\theta\sigma_x + \sin(\theta)k_\theta\sigma_y+\sin(\theta)k_r\sigma_x - \cos(\theta)k_r\sigma_y\right). \quad  \quad  \quad 
\end{eqnarray}
To simplify the problem, we neglect all the terms proportional to $\alpha_r$ as $\alpha_z$ dominates due to the strong 2DEG confinement assumed in our system. Moreover, the radial confinement creates subbands with spacing larger than those from angular momentum. Consequently, we can safely drop terms involving $k_r$, as they have a negligible contribution to the effective SO interaction within the relevant lowest-energy subbands (in any case this doesn't affect the conclusions of our work). 

After applying these simplifications, we perform the rotation into the new basis
\begin{eqnarray}
    H_{\rm SO} & = &\sum_{r,\theta,\sigma} c_{r,\theta,\sigma}^\dagger\left(\alpha_z e^{-i\sigma\theta}(-i)\frac{1}{r}\partial_\theta+\alpha_z (-i)\frac{1}{r}\partial_\theta e^{-i\sigma\theta}\right)c_{r,\theta,-\sigma}+\mathrm{h.c.}\nonumber \\
    & = & \sum_{n,l}\sum_{m,p}\sum_\sigma \frac{\alpha_z}{2}\int dr \, R_{n,l}(r)R_{m,p}(r) \int d\theta \, e^{i(p-l-\sigma)\theta}(2p-\sigma) c_{n,l,\sigma}^\dagger c_{m,p,-\sigma}+\mathrm{h.c.}\nonumber \\
    & = & \sum_{n,l}\sum_m\sum_{\sigma} \frac{\alpha_z }{l_{\omega}}\left(l+\frac{\sigma}{2}\right) \mathcal{I}_{n,m}^{l,\sigma} c_{n,l,\sigma}^\dagger c_{m,l+\sigma,-\sigma}+\mathrm{h.c.},
\end{eqnarray}
where we define the (dimensionless) radial-overlap integral
\begin{equation}
    \mathcal{I}_{n,m}^{l,\sigma} \equiv \sqrt{\frac{n!m!}{(n+l)!(m+l+\sigma)!}} \int_0^\infty dx\, x^{\left|l+\frac{\sigma}{2}\right|-\frac{1}{2}}e^{-x}\mathcal{L}_{n}^{\left| l \right|}(x) \mathcal{L}_{m}^{\left| l+\sigma \right|}(x),
\end{equation}
being $x\equiv\left(\frac{r}{l_\omega}\right)^2$. Using the Laguerre polynomial properties
\begin{eqnarray}
    \mathcal{L}_{n}^{\left| l+1 \right|}\left\{x\right\}=  \sum_{i=0}^n\mathcal{L}_{i}^{\left| l \right|}\left\{x\right\}, \quad \quad \int dx \;  x^{|l|}e^{-x}  \mathcal{L}_{m}^{\left| l \right|}\left\{x\right\}\mathcal{L}_{n}^{\left| l \right|}\left\{x\right\} = \frac{(n+l)!}{n!}\delta_{n,m},
\end{eqnarray}
it follows $\mathcal{I}_{n,m}^{l,\sigma}\sim \frac{\delta_{n,m}}{\sqrt{n+l+1}}$. This implies that coupling is (roughly) only permitted between states within the same radial subband $n=m$. And furthermore, the selection rules dictate that only angular momentum states $l$ and $l+1$ are coupled by the SO interaction. Consequently, the Hilbert space splits into decoupled subsystems labeled by the total angular momentum $J=L+\frac{1}{2}$ for each radial subband. The lowest-energy subspace corresponds to $J=\frac{1}{2}$ and is spanned by $(n,l,\sigma)=\left\{(0,0,\uparrow),(0,0,\downarrow),(0,-1,\uparrow),(0,-1,\downarrow)\right\}$. This subspace forms the basis for describing the lowest-energy physics of the system.

Hence, the SO Hamiltonian for this basis is
\begin{eqnarray}
    H_{\rm SO}=\frac{\alpha_z }{2l_{\omega}}\left(c_{0,0,\uparrow}^\dagger c_{0,1,\downarrow}-c_{0,0,\downarrow}^\dagger c_{0,-1,\uparrow}+ c_{0,1,\downarrow}^\dagger c_{0,0,\uparrow}- c_{0,-1,\uparrow}^\dagger c_{0,0,\downarrow}\right)+\mathrm{h.c.}
\end{eqnarray}
And thus the full 2DEG Hamiltonian, written in matrix form, becomes
\begin{equation}
H_{\rm 2DEG}=\begin{pmatrix}
 \tilde{H}_{\rm 2DEG} & \mathbf{0}_{\rm 4\times 4} \\
 \mathbf{0}_{\rm 4\times 4} & -\tilde{H}^*_{\rm 2DEG}
\end{pmatrix}, \quad
    \tilde{H}_{\rm 2DEG}=\begin{pmatrix}
\epsilon_{0,0} & 0 & 0 & t_{\rm SOC} \\
0 & \epsilon_{0,0} & -t_{\rm SOC} & 0\\
0 & -t_{\rm SOC} & \epsilon_{0,-1} & 0\\
t_{\rm SOC} & 0 & 0 & \epsilon_{0,1}
\end{pmatrix},
\end{equation}
with $\epsilon_{0,0}=0$, $\epsilon_{0,1}=\epsilon_{0,-1}=\frac{\hbar^2}{m^*l_\omega}$, and $t_{\rm SOC}=\frac{\alpha_z}{2l_\omega}$. This Hamiltonian has two degenerate eigenstates in the ground state: the bonding state, formed by the symmetric combination of  $\psi=(\psi_{0,0,\uparrow},\psi_{0,1,\downarrow})$; and the antibonding state, formed by the antisymmetric combination of $\psi=(\psi_{0,0,\downarrow},\psi_{0,-1,\uparrow})$. The first excited states, also degenerate in energy, are form by the opposite bonding and antibonding combinations. 

As a side note, if instead of a Rashba SO interaction we consider a Dresselhaus one, i.e., $\vec{\alpha}\cdot(\vec{\sigma}\times\vec{k}) = \alpha_z \left(\cos(\theta)k_\theta\sigma_y - \sin(\theta)k_\theta\sigma_x\right)$, then the resulting SO coupling will point towards the perpendicular direction, yielding
\begin{equation}
    \tilde{H}_{\rm 2DEG}=\begin{pmatrix}
\epsilon_{0,0} & 0 & 0 & -it_{\rm SOC} \\
0 & \epsilon_{0,0} & it_{\rm SOC} & 0\\
0 & -it_{\rm SOC} & \epsilon_{0,-1} & 0\\
it_{\rm SOC} & 0 & 0 & \epsilon_{0,1}
\end{pmatrix},
\end{equation}
which provides the same spectrum.

Performing the same transformation on the superconducting self-energy yields
\begin{eqnarray}
\Sigma_{\rm SC}^{(k)}(\omega)  = & - & \sum_{n,l}\sum_{m,p}\frac{i\Gamma_{n,m,l,p}}{\sqrt{\Delta^2-\omega^2}} \cdot  \left\{\omega\, \mathrm{sinc}\left((p-l)\delta\theta\right)\left[\cos\left((p-l)\theta_k\right)\tau_0+i\sin\left((p-l)\theta_k\right)\tau_z\right]\sigma_0 \right. \nonumber \\ 
 & & \left. + \Delta\, \mathrm{sinc}\left((p+l)\delta\theta\right)\left[\cos\left(\phi_j+(p+l)\theta_k\right)\tau_x+\sin\left(\phi_k+(p+l)\theta_k\right)\tau_y\right]\sigma_0\right\},
\end{eqnarray}
where the inter-band tunneling ratio is defined as
\begin{equation}
	\Gamma_{n,m,l,p}\equiv \Gamma_{\rm eff}\int_{l_{\rm J}}^{\infty} dr\; r R_{n,l}(r)R_{m,p}(r),
\end{equation}
with $l_{\rm J}$ the radius of the tunneling region (the radius at which the superconducting lead begins). The parameter $\theta_k$ indicates the mean angle at which the $k$-th superconducting lead is located, and $\delta\theta$ is the width of its angular coverage. This self-energy couples subbands with different angular momenta because the superconducting lead does not cover the entire circle. However, the pairing correlations are modulated by a sinc function, which suppresses couplings between subbands with significantly different angular momenta. For example, if the superconducting lead spans a quarter of the circle ($\delta\theta=\frac{\pi}{2}$), then intra-subband pairing ($l=p$) is entirely suppressed except for the $l=p=0$ case; while pairing between subbands with opposite angular momenta ($l=-p$) is maximized. Interestingly, the opposite behavior occurs for the term proportional to $\omega$. Similarly, the radial coupling term $\Gamma_{n,m,l,p}^{(k)}$ suppresses couplings between subbands with large differences in radial quantum numbers $n$ and $m$. In fact, in the limit $l_{\rm J}\rightarrow 0$ this term becomes non-zero only for $n=m$. 

This justifies that, even when considering the superconducting self-energy, it is unnecessary to include additional subbands in our lowest-energy model beyond the charge-conjugate partners. Therefore, we keep restricting our model to the lowest-energy states spanned by the quantum numbers $n=m=0$ and $\{l,p\}=\{0,\pm1\}$. In this case, the self-energy is
\begin{equation}
    \Sigma_{\rm SC}^{(k)} = \Sigma_{0,0}^{(k)} c_{0,0,\uparrow}c_{0,0,\downarrow}+ \Sigma_{0,1}^{(k)} c_{0,0,\uparrow}c_{0,1,\downarrow}+\Sigma_{0,-1}^{(k)} c_{0,0,\uparrow}c_{0,-1,\downarrow}+\Sigma_{1,1}^{(k)} c_{0,1,\uparrow}c_{0,1,\downarrow}+\Sigma_{-1,-1}^{(k)} c_{0,-1,\uparrow}c_{0,-1,\downarrow}+\mathrm{h.c.},
\end{equation}
with
\begin{eqnarray}
     \Sigma_{0,0}^{(k)} & =  & \frac{\Gamma_{0,0,0,0}}{\sqrt{\Delta^2-\omega^2}}\left[\omega\tau_0\sigma_0+\Delta\cos(\phi_k)\tau_x\sigma_0+\Delta\sin(\phi_k)\tau_y\sigma_0\right], \\
    \Sigma_{0,\pm1}^{(k)} & =  &  \frac{\Gamma_{0,0,0,\pm1}}{\sqrt{\Delta^2-\omega^2}}\,\mathrm{sinc}(\delta\theta)\left\{\omega\left[\cos\left(\theta_k\right)\tau_0\pm i\sin\left(\theta_k\right)\tau_z\right]\sigma_0 +\Delta\left[\cos\left(\phi_k\pm\theta_k\right)\tau_x+\sin\left(\phi_k\pm\theta_k\right)\tau_y\right]\sigma_0\right\}, \\
    \Sigma_{\pm1,\mp1}^{(k)} & =  &  \frac{\Gamma_{0,0,\pm1,\mp1}}{\sqrt{\Delta^2-\omega^2}}\,\left\{\mathrm{sinc}(2\delta\theta)\,\omega\left[\cos\left(2\theta_k\right)\tau_0\pm i\sin\left(2\theta_k\right)\tau_z\right]\sigma_0 +\Delta\left[\cos\left(\phi_k\right)\tau_x+\sin\left(\phi_k\right)\tau_y\right]\sigma_0\right\}, \; \\
    \Sigma_{\pm1,\pm1}^{(k)} & =  &  \frac{\Gamma_{0,0,\pm1,\pm1}}{\sqrt{\Delta^2-\omega^2}}\,\left\{\omega\tau_0\sigma_0+\mathrm{sinc}(2\delta\theta)\,\Delta\left[\cos\left(\phi_k\pm 2\theta_k\right)\tau_x+\sin\left(\phi_k\pm 2\theta_k\right)\tau_y\right]\sigma_0\right\}.
\end{eqnarray}
Notice that for $\delta\theta=\pi/2$ and $\omega\rightarrow 0$, one has
\begin{eqnarray}
    \Sigma_{0,0}^{(k)} & = & \Gamma_{0,0,0,0}\left[\cos(\phi_k)\tau_x+\sin(\phi_k)\tau_y\right]\sigma_0, \\
    \Sigma_{0,\pm1}^{(k)} & = & \Gamma_{0,0,0,\pm1}\left[\cos\left(\phi_k\pm\theta_k\right)\tau_x+\sin\left(\phi_k\pm\theta_k\right)\tau_y\right]\sigma_0, \\
    \Sigma_{\pm1,\mp1}^{(k)} & =  &\Gamma_{0,0,\pm1,\mp1}\left[\cos\left(\phi_k\right)\tau_x+\sin\left(\phi_k\right)\tau_y\right]\sigma_0, \\
    \Sigma_{\pm1,\pm1}^{(k)} & = &  0.
\end{eqnarray}
Hence, in the basis $\psi=(\psi_{0,0,\uparrow},\psi_{0,0,\downarrow},\psi_{0,-1,\uparrow},\psi_{0,1,\downarrow},-\psi^*_{0,0,\downarrow},\psi^*_{0,0,\uparrow},-\psi^*_{0,1,\downarrow},\psi^*_{0,-1,\uparrow})$ and simplifying the notation $\Gamma_{0,0,i,j}\equiv\Gamma_{i,j}$, the self-energy in this limit is
\begin{equation}
    \Sigma_{\rm SC}^{(k)}=\begin{pmatrix}
 \mathbf{0}_{\rm 4\times 4} & \tilde{\Sigma}_k \\
 \tilde{\Sigma}_k^\dagger & \mathbf{0}_{\rm 4\times 4}
\end{pmatrix}, \quad 
    \tilde{\Sigma}_k=e^{-i\phi_k}\begin{pmatrix}
\Gamma_{0,0} & 0 & \Gamma_{0,1}e^{-i\theta_k} & 0\\
0 & \Gamma_{0,0} & 0 & \Gamma_{0,1}e^{i\theta_k}\\
\Gamma_{0,1}e^{i\theta_k} & 0 & \Gamma_{1,-1} & 0\\
0 & \Gamma_{0,1}e^{-i\theta_k} & 0 & \Gamma_{1,-1}
\end{pmatrix},
\end{equation}
where we have used the properties $\Gamma_{i,j}=\Gamma_{j,i}$ and $\Gamma_{0,1}=\Gamma_{0,-1}$.

\subsection{Minimal model for a double multi-terminal Josephson junction}

We now include the coupling between two adjacent multi-terminal Josephson junctions, which we model as a quantum point contact (QPC): a very narrow constriction in the 2DEG connecting two different junctions which possesses a significantly different chemical potential. To describe two junctions, namely the left and right junctions, we double the Hilbert space and introduce an additional site to represent the QPC. The total Hamiltonian reads
\begin{equation}
    H =  H_{\rm L} + H_{\rm R} + H_{\rm QPC} + H_{\rm t},
\end{equation}
where $H_{\rm L,R}$ is the Hamiltonian derived before for a single multi-terminal Josephson junction acting on the left $c^{\dagger}_{n,l,\sigma,\mathrm{L}}$ or right $c^{\dagger}_{n,l,\sigma,\mathrm{R}}$ junction states. The Hamiltonian describing the QPC is
\begin{equation}
    H_{\rm QPC} =  \sum_{\sigma,n}\epsilon_{n,\mathrm{QPC}} c^{\dagger}_{\sigma,n,\mathrm{QPC}}c_{\sigma,n,\mathrm{QPC}} + \mathrm{h.c.},
\end{equation}
where $\epsilon_{n,\mathrm{QPC}}$ are the energy levels in the QPC region, and $c^{\dagger}_{\sigma,n,\mathrm{QPC}}$ the creator operator acting on the states of the QPC site. Notice that we are assuming that the levels in the QPC are spin-degenerate since the SO interaction in the QPC is too small to make any effect. The tunneling Hamiltonian is
\begin{equation}
    H_{\rm t}   =  \sum_n t_{\rm QPC}\begin{pmatrix}
    c^{\dagger}_{\uparrow,n,\mathrm{QPC}} & c^{\dagger}_{\downarrow,n,\mathrm{QPC}}
    \end{pmatrix} \cdot
    e^{-i\vec{\lambda}_{\rm SOC}\cdot\vec{\sigma}}\cdot\left[
    \begin{pmatrix} 
    c_{\uparrow,\mathrm{L}}(l_{\rm QPC},\theta_{\rm L})  \\
    c_{\downarrow,\mathrm{L}}(l_{\rm QPC},\theta_{\rm L})
    \end{pmatrix}+
    \begin{pmatrix}
    c_{\uparrow,\mathrm{R}}(l_{\rm QPC},\theta_{\rm R})  \\
    c_{\downarrow,\mathrm{R}}(l_{\rm QPC},\theta_{\rm R})
    \end{pmatrix}\right]+ \mathrm{h.c.}
\end{equation}
The parameter $t_{\rm QPC}$ represents the strength of hopping between the QPC and each junction, which takes the same value for both junctions for simplicity; and $\vec{\lambda}_{\rm SOC}$ represents the SO phase resulting from the SO coupling between the QPC and the junctions. As we are going to show, the results are independent of $\vec{\lambda}_{\rm SOC}$ and therefore we do not provide further details on its description. The angles $\theta_{\rm L,R}$ are the angular position of the QPC with respect to the left/right junction, and $l_{\rm QPC}$ is the distance between the QPC site and the junction's center.

Notice that, at this stage, $H_{\rm t}$ is written in terms of position, rather than the junction levels. But before performing the rotation, we first integrate out the degrees of freedom of the QPC using second-order perturbation theory, assuming that $t_{\rm QPC}\ll \epsilon_{\rm QPC}$, and also that $\epsilon_{\rm QPC}$ is off-resonant to the junction levels. This provides the effective coupling Hamiltonian
\begin{eqnarray}
    H_{\rm J} \equiv H_{\rm t}^\dagger \left(E-H_{\rm QPC}\right)^{-1} H_{\rm t} & = & \sum_{n,\sigma}\frac{|t_{\rm QPC}|^2}{E-\epsilon_{n,\mathrm{QPC}}}c^\dagger_{\sigma,\mathrm{R}}(l_{\rm QPC},\theta_{\rm R})c_{\sigma,\mathrm{L}}(l_{\rm QPC},\theta_{\rm L}) + \mathrm{h.c.} \nonumber \\
    & = &  \sum_{\sigma}\int d\omega\; \rho_{\rm QPC}(\omega)\frac{|t_{\rm QPC}|^2}{E-\omega}c^\dagger_{\sigma,\mathrm{R}}(l_{\rm QPC},\theta_{\rm R})c_{\sigma,\mathrm{L}}(l_{\rm QPC},\theta_{\rm L}) + \mathrm{h.c.} \nonumber \\
    & \simeq &  \sum_{\sigma} \pi \rho_{\rm QPC}(0)|t_{\rm QPC}|^2c^\dagger_{\sigma,\mathrm{R}}(l_{\rm QPC},\theta_{\rm R})c_{\sigma,\mathrm{L}}(l_{\rm QPC},\theta_{\rm L}) + \mathrm{h.c.},
\end{eqnarray}
where $\rho_{\rm QPC}(0)$ is the DOS of the QPC at its Fermi level. We now rotate this Hamiltonian into the same basis for each 3TJJ
\begin{equation}
    H_{\rm J}=-\sum_{\sigma}\sum_{n,l}\sum_{m,p}J_{\rm eff} e^{-i(l\theta_{\rm L}-p\theta_{\rm R})} R_{n,l}(l_{\rm QPC}) R_{m,p}(l_{\rm QPC}) c_{n,l,\sigma,\mathrm{L}}^\dagger c_{m,p,\sigma,\mathrm{R}},
\end{equation}
where $J_{\rm eff}= \pi \rho_{\rm QPC}(0)|t_{\rm QPC}|^2$ is the effective coupling between the QDs. Notice that this term couples all the angular momentum subbands and, more importantly, it is spin-preserving. This is because the QPC does not include spin-splitting effects and, as a consequence, the resulting effective coupling must be spin-preserving (in second-order perturbation theory).

\section{Numerical Methods and Simulation Details}
\label{SM:numerics}

The eigenvalues $E_\alpha$ (energies) and eigenfunctions $\Psi_\alpha$ of the minimal Hamiltonian can be determined through a self-consistent numerical diagonalization of the equation $H\Psi_\alpha=E_\alpha\Psi_\alpha$. Alternatively, the Green's function of the total system can be computed through
\begin{equation}
    \mathcal{G}(\omega)=\left(\omega\sigma_0\tau_0-H(\omega)\right)^{-1},
    \label{Eq:Greens-function}
\end{equation}
and the density of states (DOS) can be extracted from it as
\begin{equation}
    \mathrm{DOS}(\omega) = -\frac{1}{\pi}\mathrm{Im}\left\{ \mathrm{Tr}\left\{\mathcal{G}(\omega)\right\}\right\},
\end{equation}
where $\mathrm{Im}\left\{\cdot\right\}$ and $\mathrm{Tr}\left\{\cdot\right\}$ are the imaginary part and trace operations, respectively. Since the system must have even parity at the time-reversal invariant points $\phi_k=0$ and $\phi_k=2\pi$, the parity at any $\phi_k$ can be found by computing~\cite{Riwar:NatCom16}
\begin{equation}
    \mathcal{Q} = \mathrm{sign}\left\{\mathrm{Pf}\left\{H\sigma_y\tau_y\mathcal{K}\right\}\right\},
\end{equation}
where $\Theta=\sigma_y\tau_y\mathcal{K}$ represents the time-reversal operator in the chosen basis, with $\mathcal{K}$ the charge-conjugate operator.

For the full-model simulations, we use the Hamiltonian described in Eq.~\eqref{Eq:H_total}, which is discretized on a rectangular grid using finite difference methods. The continuous position $\vec{r}$
is replaced with discrete points $r_j$ in the grid, and the derivatives $\vec{k}\rightarrow-i\vec{\nabla}$ are approximated by finite differences. We use the routines implemented in Refs.~\onlinecite{Escribano:software, Escribano:tesis}. The resulting Hamiltonian is then diagonalized self-consistently to obtain the energies and eigenfunctions. Alternatively, the Green’s function of the system can be computed by directly inverting the Hamiltonian, after including the term $\omega\sigma_0\tau_0$, as in Eq.~\eqref{Eq:Greens-function}.

We compute the current going through the lead $\beta$ using the non-equilibrium Keldysh approach~\cite{Burset:tesis}
\begin{equation}
    I_{\beta} = \sum_{\gamma}\frac{e}{h}\int d\omega \mathrm{Tr}\left\{\sigma_0\tau_z\cdot\left(\tilde{\Gamma}_\beta \mathcal{G}^rF_\gamma\tilde{\Gamma}_\gamma \mathcal{G}^a - \tilde{\Gamma}_\gamma \mathcal{G}^rF_\beta\tilde{\Gamma}_\beta \mathcal{G}^a\right)\right\},
\end{equation}
where $F_\beta = \left(1+e^{\frac{\omega+eV_{\beta}\tau_z}{k_{\rm B}T}}\right)^{-1}\sigma_0$ is the Fermi-Dirac distribution for the lead $\beta$ with potential $V_\beta$, $k_{\rm B}$ the Boltzmann constant and $T$ the temperature. The tunneling ratios for each lead, $\tilde{\Gamma}_\beta$, can be found from the self-energy as $\tilde{\Gamma}_\beta=\mathrm{Im}\left\{\Sigma_\beta\right\}$, which for the superconducting leads is given by $\Sigma_\beta=\Sigma_{\rm SC}^{(\beta)}(\omega)$ of Eq.~\eqref{Eq:Sigma_app}, while for the normal leads is simply $\Sigma_\beta=i\Gamma_{\rm 0}\sigma_0\tau_0$, with $\Gamma_0$ the normal tunneling ratio. The retarded/advanced Green's function is $\mathcal{G}^{r/a}=\mathcal{G}(\omega\pm i\eta)$, with $\eta$ being negligibly small. 

The differential conductance is defined as the derivative of the current through the lead $\beta$ when the voltage in the lead $\gamma$ is varied, i.e., $G_{\beta\gamma} = \frac{dI_\beta}{dV_\gamma}$. In our setup, we distinguish two scenarios, the so-called local conductance, in which $\beta=\gamma$ and thus,
\begin{equation}
    G_{\beta\beta} = \frac{dI_\beta}{dV_\beta}=-\sum_{\gamma\neq \beta}\frac{e}{h}\int d\omega \; \mathrm{Tr}\left\{\sigma_0\tau_z\cdot\left( \tilde{\Gamma}_\gamma \mathcal{G}^rF_\beta'\tilde{\Gamma}_\beta \mathcal{G}^a\right)\right\},
\end{equation}
and the non-local conductance, in which $\beta\neq\gamma$,
\begin{equation}
    G_{\beta\gamma} = \frac{dI_\beta}{dV_\gamma}=\frac{e}{h}\int d\omega \;  \mathrm{Tr}\left\{\sigma_0\tau_z\cdot\left( \tilde{\Gamma}_\beta \mathcal{G}^rF_\gamma'\tilde{\Gamma}_\gamma \mathcal{G}^a\right)\right\},
\end{equation}
where
\begin{equation}
    F_\beta' = \frac{\partial F_\beta}{\partial V_\beta} = - \frac{e}{4k_{\rm B}T}\cdot\frac{1}{\cosh^2\left(\frac{1}{2k_{\rm B} T}\left(\omega+ eV_\beta\tau_z\right)\right)}\sigma_0\tau_z.
\end{equation}
Notice that $\frac{\partial F_\beta}{\partial V_\gamma}=0$ for all $\beta\neq\gamma$.

\section{Further Optimization of Three-Terminal Josephson Junction Parameters}
\label{SM:3TJJ}
In Sec.~\ref{Sec:2B}, we analyzed how the zero-energy modes~(ZEMs) and the energy splitting $\delta E$ are affected by some model parameters. Here, we extend this analysis to the two remaining parameters: the effective tunneling rate into the SC, $\Gamma_{\rm eff} = \Gamma_{\rm N} \delta\theta$, and the ratio of the junction size to the harmonic confinement length, $l_{\rm J}/l_{\omega}$. The results are presented in Fig.~\ref{figS1}, where we show the energy difference between the lowest and first-excited energy states, $\delta E$, vs the phases $(\phi_1,\phi_2, \phi_3)=(\phi_1,2\pi-\phi_1,0)$, and (a) $\Gamma_{\rm eff}$ or (b) $l_{\rm J}/l_{\omega}$. The green contours represents the existence of ZEMs for these values, and the black dashed lines indicate the parameters used for the rest of the simulations.  

\begin{figure}[h]
\includegraphics[width=0.5\columnwidth]{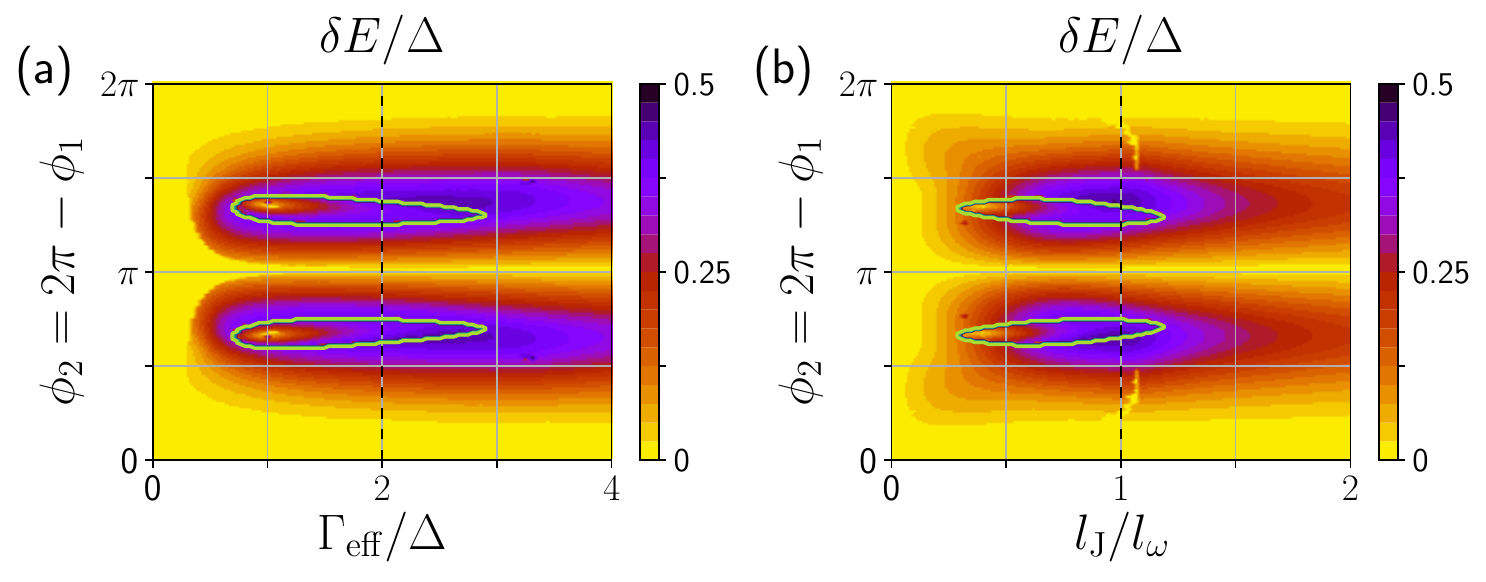}
\caption{\textbf{\label{figS1} Optimal parameters for a single 3TJJ. (a)} Energy difference between the lowest and first-excited energy states, $\delta E$ vs the phases $(\phi_1,\phi_2, \phi_3)=(\phi_1,2\pi-\phi_1,0)$ and tunneling rate into the SC, $\Gamma_{\rm eff} = \Gamma_{\rm N} \delta\theta$, normalized by the superconducting gap $\Delta$. The green contour represent the existence of the ZEMs for these values. {\textbf{(b)}} Same as in (b) but vs the ratio of the junction size to the harmonic confinement length, $l_{\rm J}/l_{\omega}$. The dashed line correspond to the same cut (and values) than the dashed line in Figs.~\ref{fig2}(c, d).}
\end{figure}

Both parameters influence the Hamiltonian similarly by modulating the strength of the coupling to the SCs. A small $\Gamma_{\rm eff}$ or $l_{\rm J}$ results in weak-coupling to the SCs, rendering the subgap states insensitive to phase variations. Conversely, a large $\Gamma_{\rm eff}$ or $l_{\rm J}$ induces strong-coupling to the SCs, pushing the Andreev bound states toward the superconducting gap and eliminating the ZEMs.  

Our findings suggest that, ideally, the harmonic confinement should be comparable to the size of the junction. This condition is (roughly) naturally satisfied due to the SCs screening the gate potentials. Furthermore, the coupling to the SCs needs to be strong, on the order of $\Delta$, and the system seems to remain robust against moderate variations in these parameters.

\section{Phase-diagram of a Double Three-terminal Josephson Junction}
\label{SM:double-3TJJ}
To provide a deeper understanding of the full parameter space in the double 3TJJ system, Fig.~\ref{figS2}(b) displays the MP (colorbar) and the zero-energy modes (ZEMs, green contours) as functions of the phases in the left 3TJJ, $(\phi_1, 2\pi - \phi_1,0)$, and the right 3TJJ, $(\tilde{\phi}_1+\delta\phi, 2\pi - \tilde{\phi}_1-\delta\phi,0)$. Different subplots correspond to different values of $\delta\phi$ and $J_{\rm eff}$, marked in Fig.~\ref{figS2}(a) (note that the markers follow the same spatial ordering).  

Figure~\ref{figS2}(a), which is the same as in Fig.~\ref{fig4}(a), presents the maximum MP across all ZEMs for all possible phase configurations, for different $\delta\phi$ and $J_{\rm eff}$. The gray area in Fig.~\ref{figS2}(a) indicates parameter regions where no ZEMs are found along the phase path $\phi_1 = 2\pi - \tilde{\phi}_1-\delta\phi$.

\begin{figure}
\includegraphics[width=0.98\columnwidth]{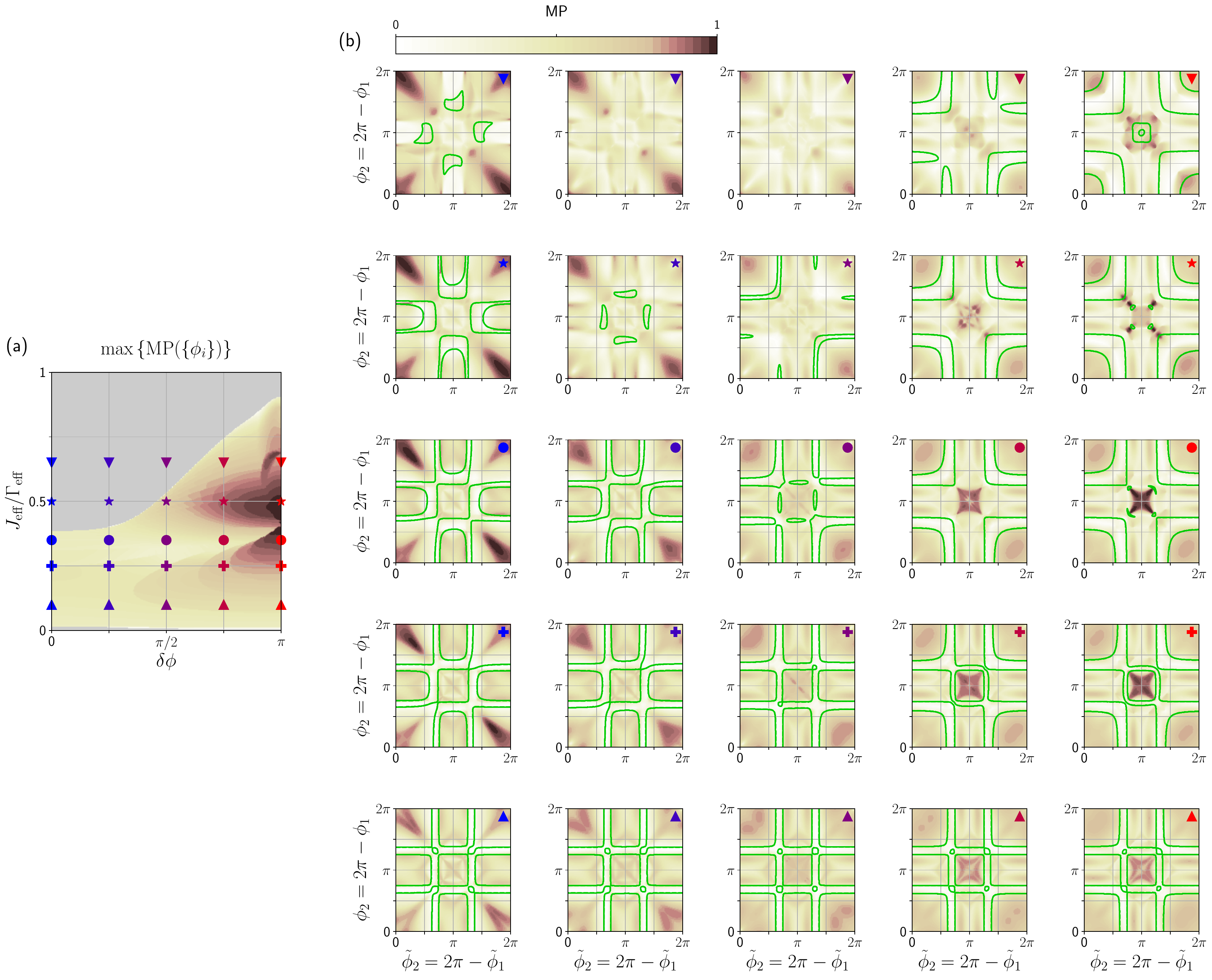}
\caption{\textbf{\label{figS2} Phase space of a double 3TJJ. (a)} Same as in Fig.~\ref{fig4}(a). \textbf{(b)} MP vs the phases in both 3TJJs, $(\phi_1, 2\pi - \phi_1,0)$ and $(\tilde{\phi}_1+\delta\phi, 2\pi - \tilde{\phi}_1-\delta\phi,0)$. Each subplot, tagged with a marker, corresponds to a simulation performed with different values  of $\delta\phi$ and $J_{\rm eff}$ marked in Fig.~\ref{figS2}(a).}
\end{figure}

\section{Poor's Man Majorana Wavefunction}
\label{SM:Continuum}

In Fig.~\ref{figS3} we show the probability density of the lowest-energy modes identified as a poor's man Majorana mode in Figs.~\ref{fig5}(c-f) (green circle). We show both states states degenerate at zero energy, with (a) even or (b) odd symmetry. Notice they are not completely symmetric because (i) they are not perfectly degenerate at zero-energy, and (ii) the system is not symmetric due to disorder and the phase configuration. We compute the MP for these modes obtaining $\mathrm{MP}\simeq0.95$.

\begin{figure}
	\begin{centering}
		\includegraphics[width=0.5\columnwidth]{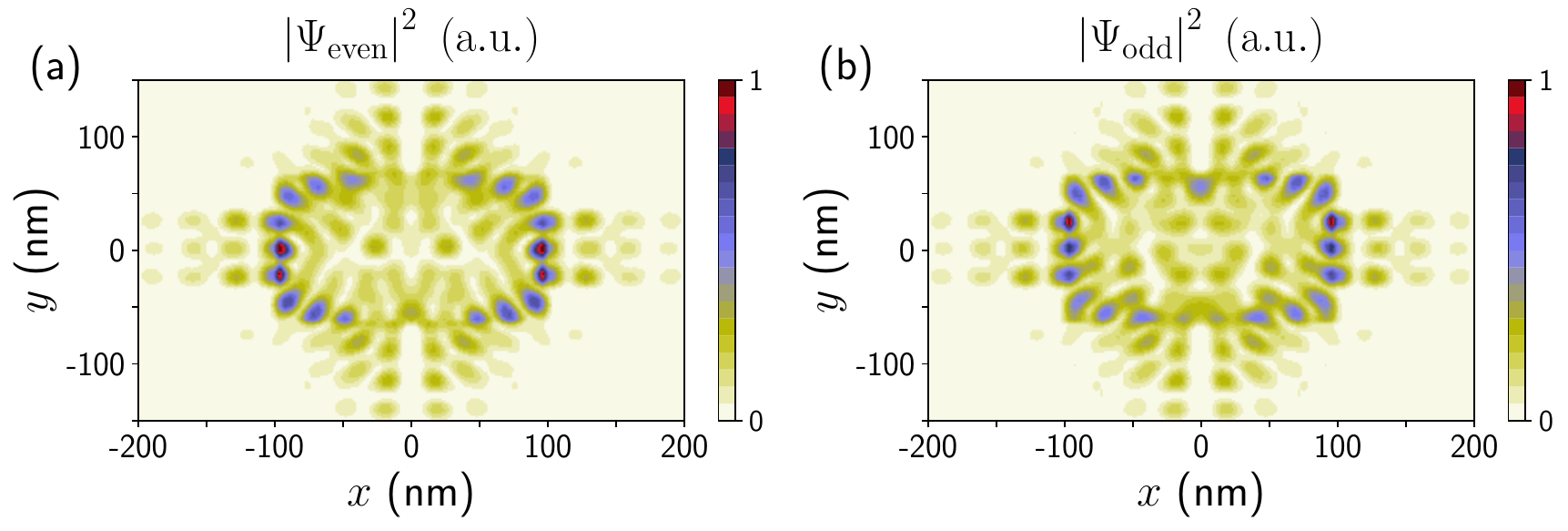}
		\par\end{centering}
	\caption{\textbf{\label{figS3} Poor's man Majorana mode wavefunction.} Probability density of the zero-energy mode identified as a poor's man Majorana mode in Fig.~\ref{fig5}(c-f) (green circle). We show both states degenerate at zero energy, with (a) even or (b) odd symmetry.  }
\end{figure}

\end{document}